\documentclass{egpubl}
\usepackage{eg2026}
\ConferencePaper        
\CGFStandardLicense

\usepackage[T1]{fontenc}
\usepackage{dfadobe}  
\usepackage{dfadobe}  
\usepackage{booktabs} 
\usepackage{xcolor}
\usepackage{subcaption}
\usepackage{amsmath}
\usepackage{amsfonts}
\usepackage{enumitem}
\usepackage[skip=2pt]{caption}
\usepackage{pgffor} 
\usepackage{cite}  
\bibtexVersion 
\BibtexOrBiblatex
\electronicVersion
\PrintedOrElectronic
\ifpdf \usepackage[pdftex]{graphicx} \pdfcompresslevel=9
\else \usepackage[dvips]{graphicx} \fi

\usepackage{egweblnk} 
\usepackage{lineno}
\nolinenumbers

\title[ProcTex: Consistent and Interactive Text-to-texture Synthesis for Part-based Procedural Models]%
      {ProcTex: Consistent and Interactive Text-to-texture Synthesis for Part-based Procedural Models}

\author{
    \large{
    Ruiqi Xu\textsuperscript{1}\thanks{Equal contribution} \hspace{5mm}
    Zihan Zhu\textsuperscript{1}\footnotemark[1] \hspace{5mm}
    Ben Ahlbrand\textsuperscript{1} \hspace{5mm}
    Srinath Sridhar\textsuperscript{1} \hspace{5mm}
    Daniel Ritchie\textsuperscript{1}
    }\\
    \large{\textsuperscript{1}Brown University}
}

\begin{document}
\pagestyle{plain}
\teaser{
 \includegraphics[width=0.9\linewidth]{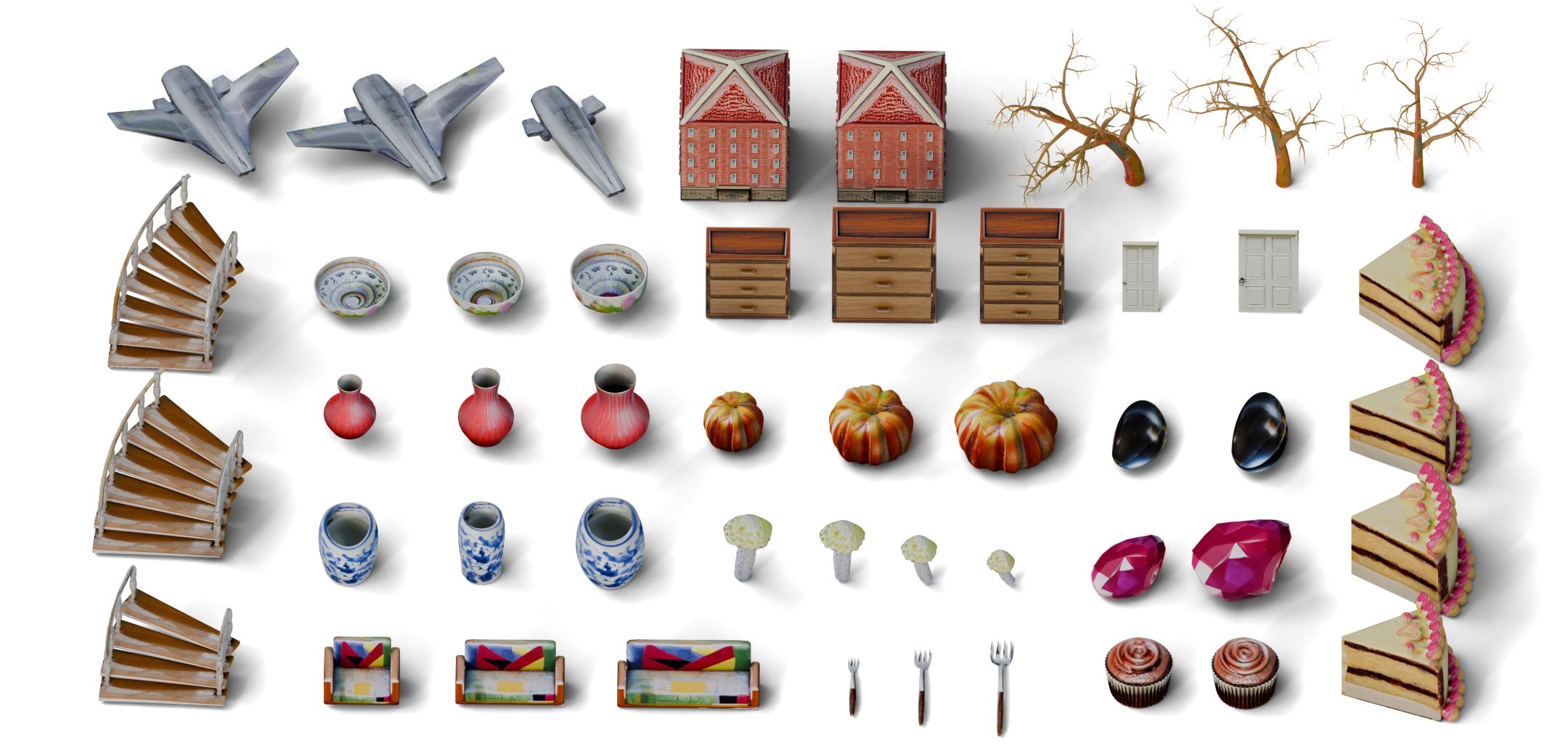}
 \centering
  \caption{Texturing results. ProcTex is a system for real-time text-guided texture synthesis on part-based procedural models. Even though small parameter changes in procedural models can induce large topological differences, ProcTex achieves consistent, high-quality part-level textures across diverse shape variations.}
\label{fig:teaser}
}

\maketitle
\begin{abstract}
   Recent advances in generative modeling have driven significant progress in text-guided texture synthesis. However, current methods focus on synthesizing texture for single static 3D object, and struggle to handle entire families of shapes, such as those produced by procedural programs. Applying existing methods naively to each procedural shape is too slow to support exploring different parameter configurations at interactive rates, and also results in inconsistent textures across the procedural shapes. To this end, we introduce ProcTex, the first text-to-texture system designed for part-based procedural models. ProcTex enables consistent and real-time text-guided texture synthesis for families of shapes, which integrates seamlessly with the interactive design flow of procedural modeling. To ensure consistency, our core approach is to synthesize texture for a template shape from the procedural model, followed by a texture transfer stage to apply the texture to other procedural shapes via solving dense correspondence. To ensure interactiveness, we propose a novel correspondence network and show that dense correspondence can be effectively learned by a neural network for procedural models. We also develop several techniques, including a retexturing pipeline to support structural variation from procedural parameters, and part-level UV texture map generation for local appearance editing. Extensive experiments on a diverse set of procedural models validate ProcTex’s ability to produce high-quality, visually consistent textures while supporting interactive applications.


\end{abstract}  

\clearpage
\section{Introduction}
In 3D content creation, texture and geometry are the two cornerstones that define how assets look and feel. While texture creation has traditionally required significant manual effort, recent advances in generative modeling allow users to create high-quality texture directly from a simple text prompt \cite{richardson2023texture, chen2023text2tex, cao2023texfusion, chen2023fantasia3d, hunyuan3d2025hunyuan3d, xiang2024structured, li2025triposg, lin2025kiss3dgen}. Meanwhile, procedural modeling has been a long-established methodology for geometry creation, which allows users to generate endless families of complex shapes by manipulating procedural parameters\cite{palubicki2009self, nishida2016interactive, raistrick2023infinite, infinigen2024indoors}. Uniting these two directions so that procedural shapes update with consistent, text-guided textures in real time would serve as a key step toward efficient 3D content creation. However, combining these two paradigms effectively remains challenging and under-explored. Existing text-guided texturing systems struggle to preserve consistency across a family of generated shapes, and often fail to produce reliable textures at interactive rates to support the design workflow that procedural modeling is built for.

\begin{figure}[t]
\centering
    \centering
    \includegraphics[width=\linewidth]{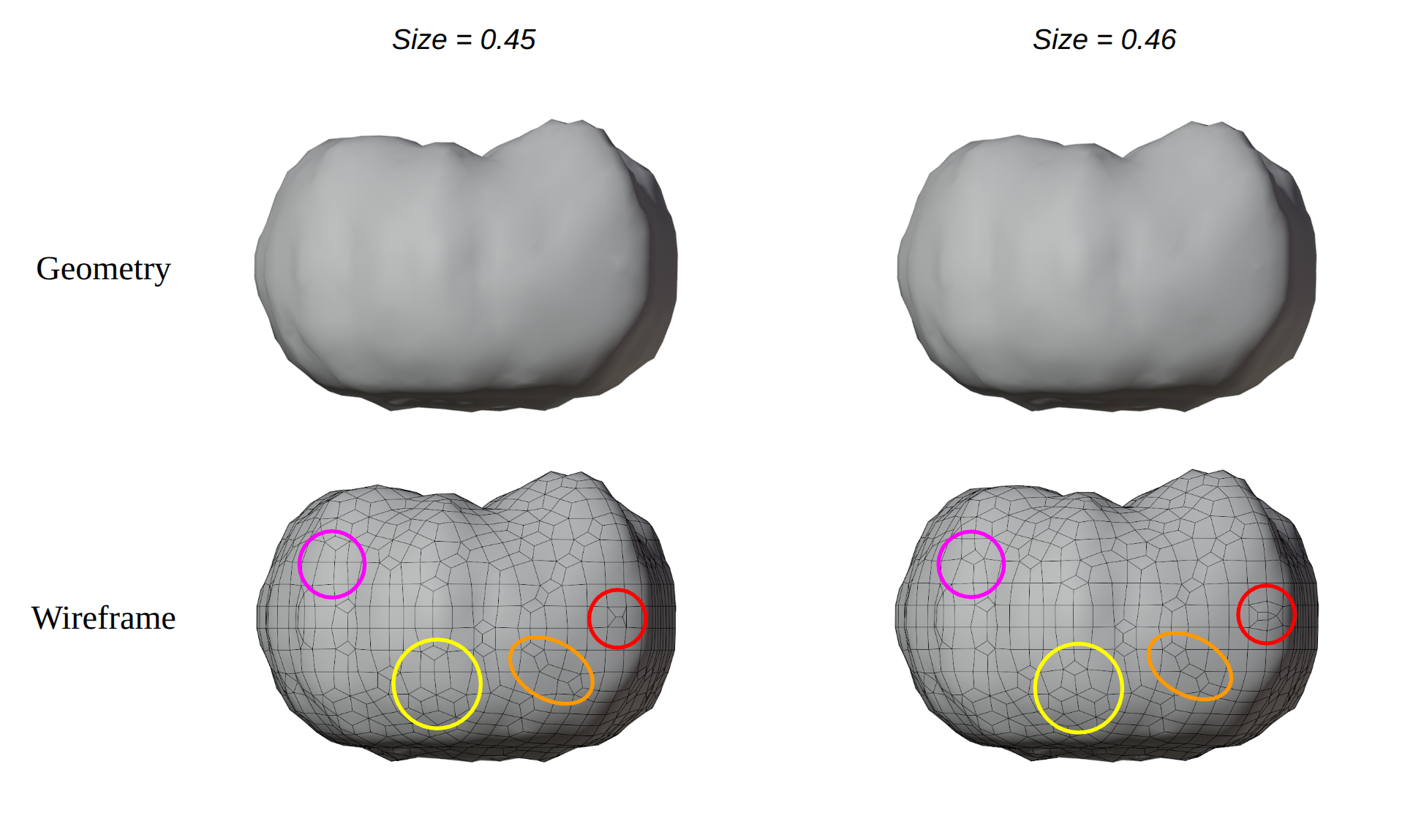}
    \caption{A small parameter change in the procedural model can result in complex topology differences, despite near-identical appearance. This figure shows an example of a pumpkin procedural model from Blender Market. The colored circles highlight connectivity changes on corresponding regions. Left: 1,518 vertices / 3,062 edges / 1,546 faces. Right: 1,491 vertices / 3,002 edges / 1,513 faces.}
    \label{fig:topology_difference}
\end{figure}

This paper presents ProcTex, a system that enables consistent and real-time text-to-texture generation for part-based procedural models. ProcTex is designed to remain agnostic to the internal workings of the procedural model and treat it as a "black box". This allows ProcTex to generalize across a wide range of generators without requiring any modification to the generator itself. We focus on the setting of part-based models, where shapes are composed of individual components (e.g., legs, arms, handles), each exhibiting bounded but meaningful geometric variation. Our key design is to frame the multi-shape texturing problem as a one-time generation followed by a learned, real-time texture transfer. ProcTex first synthesizes texture with a template procedural shape using an existing text-to-texture method, and then transfers the texture to other procedural variations. 
A critical challenge is the topological instability inherent in many real-life procedural models. As illustrated in Fig.~\ref{fig:topology_difference}, even a small parameter tweak with subtle visual change can lead to a complex modification to the underlying mesh connectivity. Since ProcTex makes no assumption about the internal rules of procedural models, such topology change is not analytically tractable. To handle these cases, ProcTex improves on recent cage-based deformation methods \cite{yifan2020neural} to solve for efficient and robust texture mapping by estimating dense correspondence.

To achieve real-time texture transfer for interactive workflows, we introduce ProcCorrNet, a novel correspondence network trained to predict dense surface correspondence for procedural shapes. ProcTex leverages cage-based solutions from the previous step to train ProcCorrNet, and then applies the learned correspondences at runtime with a single feed-forward pass for arbitrary shapes produced by the procedural model. Further, ProcTex supports part-based procedural models with both continuous parametric variations and structural changes such as part addition and deletion. Additionally, ProcTex generates textures as part-level UV-mapped images, which enables convenient part-level texture editing for users.

We demonstrate the effectiveness of our model by evaluating it on a diverse set of commercial and open-source part-based procedural models spanning many object categories and varying degrees of geometric deformations. Both qualitative and quantitative results show that ProcTex produces higher-quality, more consistent textures than baselines, while operating at interactive rates. To the best of our knowledge, no prior work has studied text-guided texture generation for procedural modeling. Our results demonstrate the potential of combining generative texture synthesis with procedural modeling to facilitate efficient 3D content creation.

To summarize the main contributions of this work: 
\begin{enumerate}[label=(\arabic*), itemsep=2pt, topsep=4pt]
\item We propose ProcTex, a system that enables text-guided texturing for part-based procedural models. Our system supports procedural models with structural and continuous variations, generates consistent textures, supports interactive parameter manipulation, and synthesizes UV texture maps that can be conveniently edited.

\item We introduce ProcCorrNet, a novel correspondence network as part of our ProcTex system, which effectively estimates dense shape correspondence, to facilitate texture synthesis for procedural models at interactive rates.

\item We improve recent cage-based deformation methods to robustly estimate dense correspondence estimation for part-based procedural models.

\end{enumerate}

\vspace{-5pt}
\section{Related Work}

\subsection{Procedural Modeling}
    Procedural modeling is the process of using programs or rules to produce visual content. It has a long history in computer graphics \cite{sutherland1964sketch} and is widely used in areas such as 3D geometry modeling \cite{nishida2016interactive, palubicki2009self, merrell2010computer}, material and texture design \cite{guerrero2022matformer, perlin1985image, worley1996cellular}, and fabrication \cite{ghali2008constructive, stroud2006boundary}.

    One well-known technique is \textit{L-systems}, which employ context-free string rewriting systems to generate branching structures, such as organic objects \cite{prusinkiewicz1986graphical, prusinkiewicz1988development, prusinkiewicz1996systems}. Another popular class is \textit{shape grammars}, which use shape rules and an engine that applies these rules to create assets such as facades \cite{muller2007image, martinovic2013bayesian}, buildings \cite{muller2006procedural}, and cities \cite{kelly2021cityengine}. More recently, researchers have explored combining procedural modeling with deep learning for 3D shape modeling, with applications in inferring 3D shape programs \cite{ellis2019write, xu2021inferring, deng2022unsupervised}, 3D shape generation \cite{pearl2025geocode, nishida2018procedural, jones2020shapeassembly, wu2021deepcad, zhang2024scene}, and parametric shape editing \cite{kim2024meshup, ganeshan2024parsel}. 
    

    This paper aims to enable text-guided texture synthesis for 3D part-based procedural models. These models are typically defined by interfaces that map program parameters to deformable 3D components, allowing users to randomize the parameter inputs and interactively explore the design space of the procedural models. Our model leverages this interface to support efficient, real-time texture generation. Importantly, our method makes minimal assumptions about the procedural models themselves and require no access to procedural rules or priors. This generality ensures our model is agnostic to object categories, specific generator architectures, and the dimensionality of the parameter space.
\subsection{Text-guided texture generation systems}
    Inspired the recent success of 2D image generation using text-guided diffusion models, several works attempt to leverage pretrained 2D diffusion models \cite{rombach2022high, saharia2022photorealistic} to guide texture synthesis for 3D shapes. A popular line of work \cite{poole2022dreamfusion, lorraine2023att3d, qian2023magic123,
    metzer2023latent,
    lin2023magic3d, shi2023MVDream, qiu2024richdreamer} is to represent the 3D object as a neural radiance field \cite{mildenhall2021nerf} and optimize it through a score distillation sampling objective and its variants\cite{poole2022dreamfusion, wang2024prolificdreamer, wang2023score}. Researchers have also explored training Gaussian Splatting models \cite{kerbl3Dgaussians} via supervision from pretrained 2D diffusion models \cite{tang2023dreamgaussian, yi2023gaussiandreamer}. While these methods achieve impressive results, extracting explicit 3D assets (e.g., meshes) from such implicit representations remains challenging, which limits their applicability to be integrated into standard computer graphics workflows.

    Aside from implicit representations, A major line of work focuses on generating textured 3D meshes under text guidance. Early methods such as \cite{khalid2022clipmesh} and \cite{text2mesh} use CLIP-based signals \cite{clip} to optimize textures that align with a given textual description. More recently, \cite{chen2023fantasia3d, deng2024flashtex, huo2024texgen, wei2023taps3d, liu2024text, xiang2024structured, li2025triposg} leverage GAN \cite{goodfellow2014generative} and rectified flow \cite{albergo2022building, lipman2022flow} models to optimize textures and optionally the geometries with text guidance. There are also several works that represent textures in UV-mapped images and directly generate these texture images \cite{tang2024intex, cao2023texfusion, gao2024genesistexadaptingimagedenoising, yeh2024texturedreamer, lin2025kiss3dgen, hunyuan3d2025hunyuan3d}. These methods improves the editability of resulting assets as well as avoids the lengthy SDS optimization. However, these approaches typically process only a single mesh at a time. Re-running the training for each new shape is prohibitively expensive in interactive settings, and there is no inherent guarantee of consistency across textures generated independently by these models.

    Meanwhile, another line of research aims to develop generalizable 3D generative models \cite{yu2024texgen, bensadoun2024meta3dtexturegenfast, liu2023zero1to3, hong2024lrmlargereconstructionmodel}. These systems are trained on large-scale 3D datasets and can synthesize arbitrary textured 3D outputs quickly during inference. However, such approaches do not accept a family of shapes as input, which limits their use in the context of procedural modeling. To the best of our knowledge, existing works that specifically address text-driven texture generation for multiple shapes \cite{dong2024coin3d} are limited to simple primitive forms and do not readily scale to complex or large families of geometries. In contrast, our work bridges this gap by enabling text-guided texture generation for procedurally generated shapes, while simultaneously ensuring texture consistency and supporting real-time inference.




\section{Method}
\begin{figure*}[t]
\centering
    \centering
    \includegraphics[width=\linewidth]{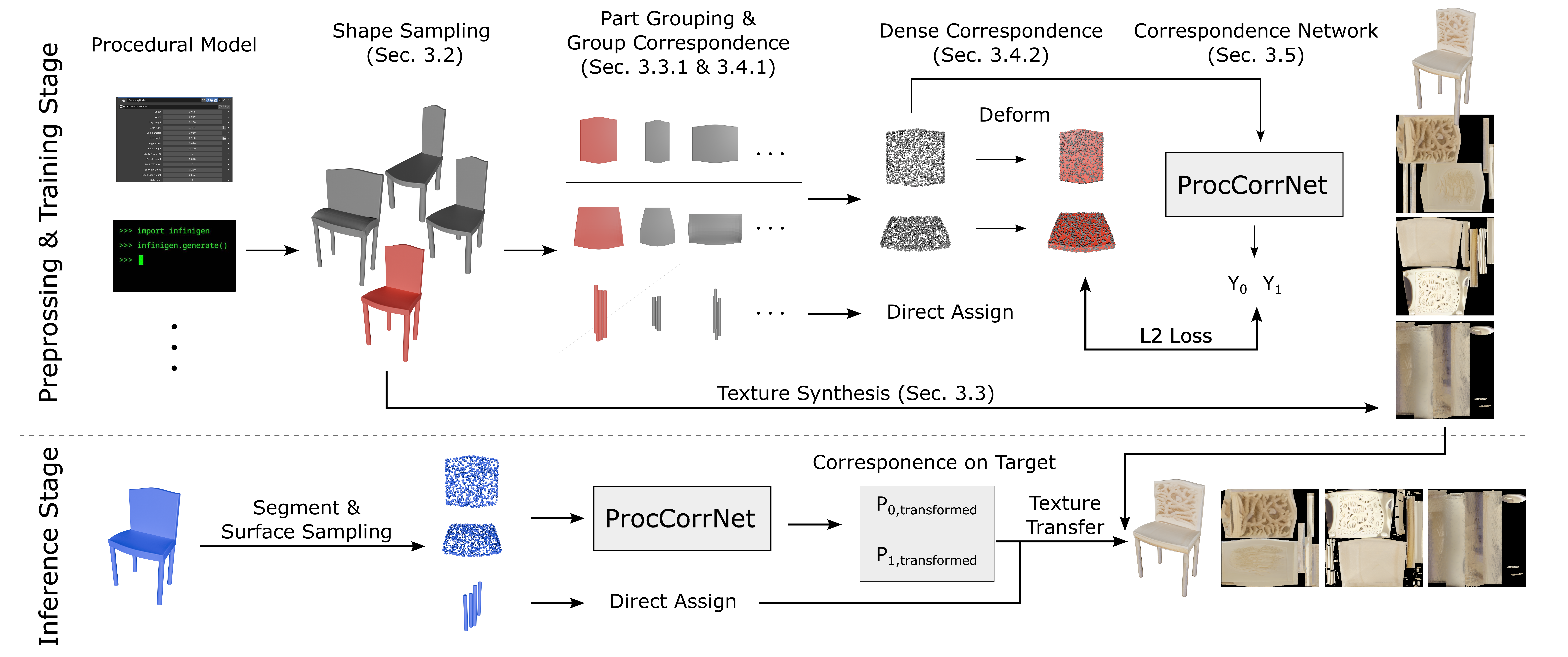}
    \caption{\textbf{ProcTex Overview}. ProcTex consists of three modules: (1) a texture synthesis pipeline to generate part-level UV texture images; (2) a shape matching module to establish dense surface correspondence across the sampled shapes; (3) a correspondence network that predicts the corresponding points on the target surface for given points on the source surface. ProcTex is trained on a sampled set of meshes during the pre-processing stage. For inference, ProcTex predicts the corresponding points and perform texture transfer any procedural mesh in real-time.}
    \label{fig:model_overview}
\end{figure*}

\subsection{Problem Statement and Approach}
The goal of this paper is to develop a system to enable text-based
texture synthesis for part-based procedural models. Our system must generate textures that are consistent across geometries, and also produce textures at interactive rates to support real-time exploration. The key design of our system is to generate UV texture images for a template mesh sampled from the procedural model, and then train a neural network to perform learned, real-time texture transfer for all other procedural meshes. Fig. \ref{fig:model_overview} provides an overview of our system.


Our model decouples the task into two stages. In the pre-processing stage, ProcTex starts by randomly sampling meshes and identifying one template shape from samples. ProcTex then builds on an existing texture inpainting method to generate part-level UV texture images for the template shape. Next, ProcTex estimates dense surface correspondence between the sampled meshes and the template mesh through solving cage-based deformation \cite{yifan2020neural}. The UV textures are then transferred to other procedural meshes according to the dense surface correspondence. Then, ProcTex trains a correspondence network with the goal of learning dense correspondence for the entire family of shapes produced by the procedural model. During the inference stage, the learned correspondence network predicts the corresponding dense correspondence with just one single feedforward process and effectively transfers the generated texture to the procedural shapes at interactive rates.

\subsection{Parameter sampling and selecting template mesh}

The first step of our model is to sample meshes $M_{1, ..., n}$ that span the parameter space of the procedural model. The key reason for this step is to capture the full range of configurations and ensure that ProcTex can handle any procedural setting at inference. To achieve this goal, we systematically sample both continuous and discrete parameters of the procedural model. Continuous parameters typically control the size, dimensions, or locations of a shape, while discrete parameters, on the other hand, control the presence and quantity of certain geometry components in the object. For each parameter of interest, we determine lower and upper bounds within which the model produces plausible shapes for the object category. We then generate a collection of meshes by randomly sampling values within these parameter intervals from the procedural model. 

After sampling shapes from the procedural models that span the design space, the next step is to select a template shape for texture generation. While any sampled mesh can be used in principle, we compute the mean of all sampled parameter vectors, and then select the mesh whose parameters are nearest to this mean vector under Euclidean distance. This choice provides a representative template that balances across the model’s variations, and we empirically find that it improves the stability of training the correspondence network.


\subsection{Synthesizing part-level textures for template mesh}  \label{texture_generation}
To support convenient texture editing, ProcTex synthesizes textures in the form of UV-mapped texture images for each part of the template mesh. Although part segmentation for 3D shapes is generally challenging, we propose a simple yet effective approach under the context of part-based procedural models. We detail our method for part identification and grouping in Sec. \ref{part_identification_grouping}, and the texture generation pipeline in Sec. \ref{intex_modification}.

\subsubsection{Part Identification \& Grouping} \label{part_identification_grouping}
Part segmentation of a mesh is a fundamental problem in computer graphics \cite{liu2024part123, tang2024segmentmeshzeroshotmesh, zhou2023partslipenhancinglowshot3d, abdelreheem2023SATR}. While part segmentation is challenging for arbitrary shapes without any prior knowledge, our key observation is that many part-based procedural models build complex geometries by first constructing simpler components and then assembling them to form the whole shape. For example, Blender's Geometry Nodes \cite{blender} model individual components that are linked together in a tree-like structure, and Infinigen \cite{raistrick2023infinite} follows a similar paradigm for procedural asset generation. Notably, part-based procedural models tend to avoid post-hoc connectivity, as they do not explicitly add edges between components after each is generated. As a result, we can segment a mesh into meaningful components based on connectivity, which has proven effective across all the procedural models we experimented. Specifically, ProcTex runs a standard depth-first search (DFS) algorithm based on the faces of a given mesh $M_{i}$, and treats each connected component as a separate part $C_{1, ..., k}$, where $k$ denotes the number of connected components.

When a single component is repeated multiple times (e.g., two armrests of a sofa), sharing a texture map across instances reduces redundancy and prevents visual inconsistencies. ProcTex divides the components into groups $G_{1, ..., m}$ and synthesizes a shared texture image $I_{1, ..., m}$ for each group. ProcTex first group components by matching mesh topologies, which handles cases when components are direct duplication from each other. When exact topological matches are not present, ProcTex performs rigid alignment using Iterative Closest Point (ICP) \cite{ICP} among components and measures their geometric similarity with Chamfer distance. A predefined threshold is then applied to determine whether components belong to the same group. This works well because the grouped components need to preserve visual appearance, although being topologically different. Such topological variations are often introduced by mirroring or flipping operations, which are commonly used to model natural symmetries in procedural generator (e.g., the wings of a spaceship). To improve robustness in such cases, ProcTex handles this by first detecting $M_{i}$'s global symmetry planes along the x, y, and z axes. If a candidate component pair lies on opposite sides of a detected symmetry axis, one component is reflected before applying ICP alignment.

\subsubsection{Texture generation} \label{intex_modification}
We build on InTex \cite{tang2024intex} pipeline to serve as the text-guided texturing module for the template mesh $M_{T}$. InTex iteratively renders the mesh from a set of camera poses and obtains per-view RGB images, inpainting masks, and depth maps. A depth-aware inpainting diffusion model then synthesizes an inpainted image conditioned on the text prompt. The result is back-projected onto the UV-mapped texture images. We refer readers to the original InTex paper for additional details.

We extend InTex to operate at the component-group level defined in Sec. \ref{part_identification_grouping}. Let $G_{1, ..., m}$ be the component groups and $I_{1, ..., m}$ their associated texture images. For back-projection, ProcTex maintains a per-component depth map and compares it with the global depth map, assigning each pixel to the nearest component before projecting the inpainted result onto the corresponding $I_{j}$. This strategy decomposes the global texture image into component-level maps, while preventing mutual overwrite among overlapping components.

We favor an inpainting-based approach because it is significantly faster and achieves similar visual quality compared to optimization-based methods \cite{poole2022dreamfusion, chen2023fantasia3d, liu2024text}. Nonetheless, ProcTex is independent of the specific texturing algorithm. Any single-object texturing pipeline that outputs UV images (e.g., \cite{richardson2023texture, chen2023text2tex}) can replace InTex with minimal integration effort, which ensures that ProcTex is compatible with future advances in text-to-texture methods.

\begin{figure*}[t!]
\centering
    \centering
    \includegraphics[width=\linewidth]{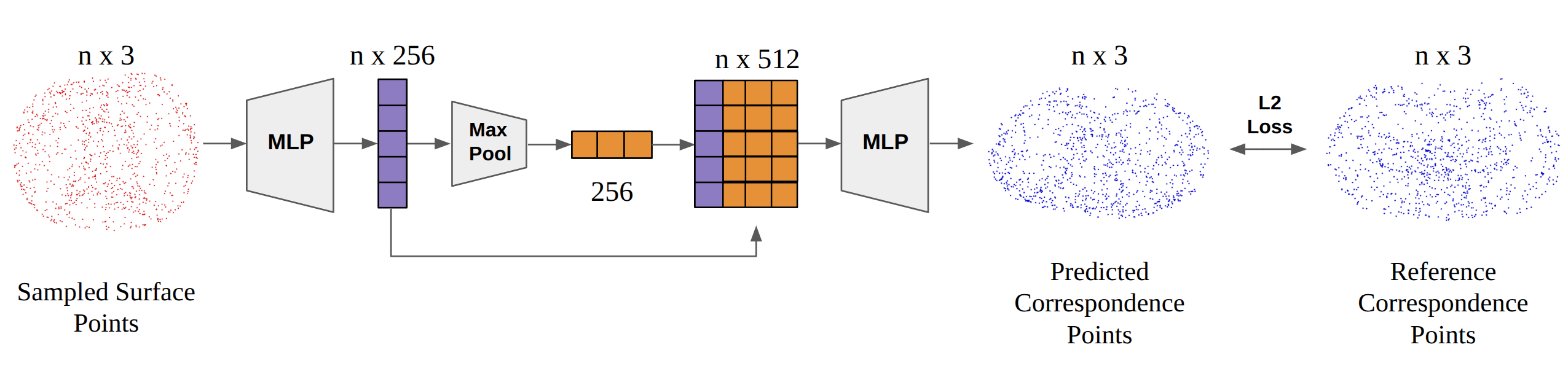}
    \caption{\textbf{ProcCorrNet Architecture}. Given input 3D points from a procedural component, local features are extracted via a 4-layer MLP. A global descriptor is computed by max pooling and concatenated with local features to form a joint representation. A second 4-layer MLP predicts per-point correspondences in 3D space. The network is trained with supervision from cage-deformed correspondences on the template mesh (Sec. \ref{dense_correspondence}) using L2 loss.}
    \label{fig:proccorrnet_architecture}
\end{figure*}

\subsection{Transferring textures for procedural meshes} \label{texture_transfer}
After synthesizing part-level textures of the template mesh $M_{T}$, ProcTex transfers the textures onto other shapes. This approach ensures that all shapes reference the same texture images and therefore visual consistency is guaranteed across the procedural shapes. 
For texture transfer, our key design is to build a dense surface correspondence between the template mesh $M_{T}$ and other meshes $M_{i}, i\in\{1, ..., n\}$, after which the correspondence is used to bake texture images for $M_i$ that follow each mesh's own UV coordinates. For shapes with a single connected component, one shape is treated as a whole to calculate surface correspondence. For shapes with multiple component groups, the surface correspondence is performed for each component group independently to apply texture transfer. 

\subsubsection{Identifying group correspondence} \label{correspondence_across_shapes}
    Prior to establishing dense surface correspondence, ProcTex first determines which component groups of a sampled mesh $M_i$ correspond to those of $M_T$. Sec. \ref{part_identification_grouping} yields two sets of their component groups $G_{i1}, ..., G_{in}$ and $G_{T1}, ..., G_{Tn}$. The order of the groups are ruled by when a component is visited by DFS and it is not guaranteed that $G_{ij}$ corresponds to $G_{Tj}, \forall j \in 1, ..., n$. 
    To match groups, ProcTex performs rigid ICP alignment for every pair $(G_{ij}, G_{Tk})$. For each alignment, we compute the mean Chamfer distance and add a rotation penalty that discourages large rotations. This regularization prevents geometrically similar—but semantically distinct—components (e.g., horizontal vs.\ vertical cuboids in furniture) from being matched solely due to a small reduction in Chamfer distance. For every $G_{ij}$, the final assignment is obtained by selecting the $G_{Tk}$ that minimizes the regularized distance.
    
    

\subsubsection{Solving dense correspondence} \label{dense_correspondence}
After groups are corresponded, ProcTex establishes dense correspondence between reference components. For each group, let $C_s$ and $C_t$ denote the reference component in the template mesh $M_t$ and the corresponding reference component in a sampled mesh $M_i$. ProcTex first checks the mesh topology of $C_s$ and $C_t$. In scenarios when this holds (e.g., components are direct duplicates of each other), ProcTex directly assign the UVs from $C_s$ to all components that belong to the same group as $C_t$.

When topology consistency does not hold, establishing dense correspondence between components becomes a classic shape matching problem, which is a well-studied topic in geometry processing \cite{litany2017deepfunctionalmapsstructured, cao2023unsupervised, sharma2020weaklysuperviseddeepfunctional, eisenberger2019smoothshellsmultiscaleshape, yifan2020neural}. In ProcTex, we modify existing caged-based deformation methods \cite{yifan2020neural}. Our modified version achieves faster convergence and significantly improves performance on thin structures, which are essential for tableware procedural models. We describe our improved optimization objective below.

\begin{equation}
    \mathcal{L} = \alpha_{\text{mvc}}\mathcal{L}_{\text{mvc}} + \alpha_{\text{normal}}\mathcal{L}_{\text{normal}} + \mathcal{L}_{\text{align}}  + \mathcal{L}_\text{surface}
    \label{cage_loss}
\end{equation}

We first align the source and target mesh point clouds using Kabsch-Umeyama ICP \cite{umeyamaicp} to get a best fit for scale, rotation and translation.
We optimize a single batch of pairs, this contrasts with \cite{yifan2020neural} where they train a deformation network over many shapes.
This non-rigid deformation is parameterized by a cage with mean-value coordinate weights \cite{mvc}, initialized by a subdivided axis-aligned bounding box (AABB) in contrast to the UV sphere used in \cite{yifan2020neural}.
Given we are solving a single pair shape matching problem, we modify their normal loss and use ground truth barycentric interpolated vertex normals from the initial mesh to pair ground truth position + normals of the target shape.
We then penalize angular distance between ground truth and SVD normals of the deformed samples (setting neighborhood size to 1) -- this greatly improves results on densely sampled surfaces with thin features where neighbors in the point cloud are ambiguous.
We also allow for arbitrary axes of symmetry (but in practice optionally enable along $z = 0$, $y = 0$ or $x = 0$) for $\mathcal{L}_{symm}$ in cases where additional fine-tuning may be desirable to improve the quality of the match (vases for example often have at least 2 axes of symmetry to exploit).
We found that introducing a surface penalty ( $\mathcal{L}_{surface}$ ), via Pytorch3D \lowercase{\texttt{point\_mesh\_face\_distance}} operator improves the tightness of the fit between shapes due to chamfer distance optimizing for the entire distribution matching but not necessarily exact matches. 

\subsubsection{Texture Baking}
\label{texture_baking}
After establishing dense surface correspondences between $C_s$ and $C_t$, we transfer texture by resampling from $C_t$ onto the UV space of $C_s$. Specifically, we uniformly sample a grid of UV coordinates in $C_s$'s UV domain, map each to a 3D point on $C_s$ via barycentric interpolation, and use the correspondence to locate the matched surface point on $C_t$. The UV coordinates of this point on $C_t$ are then used to retrieve colors from $C_t$'s texture image, which are assigned back to the original UV sample on $C_s$. The result is a baked texture aligned with $C_s$'s UV layout. This resampling step follows the standard texture transfer practice (e.g., \cite{decatur2023paintbrush}) and, in our implementation, introduces no noticeable degradation while remaining efficient enough to preserve the interactive performance of ProcTex.

\subsubsection{Re-texturing for part addition and dis-occlusion} \label{part_addition_removal}
When new components appear or previously hidden areas become exposed, the template textures may not fully cover a new procedural shape. To tackle these two challenges, we develop a re-texturing pipeline with ProcTex, which is discussed below.

For component addition, ProcTex first aligns each added component to existing ones via ICP. If a match is found, the corresponding texture map is reused. Otherwise, ProcTex re-runs the texturing pipeline in Sec. \ref{intex_modification} to synthesize new textures for the added component. The textures for existing components are not updated in order to maintain visual consistency with the template mesh.

Structural variations such as removal, scaling, displacement can reveal texels that were unseen in the template mesh and thus render black. ProcTex tracks an “updated” mask for each texture map. Texels marked as previously filled remain frozen, and only the newly exposed pixels are synthesized. This design ensures the re-texturing process selectively updates only the necessary areas, maintaining texture consistency among shapes with structural changes.

\begin{figure}[t]
    \includegraphics[width=\linewidth]{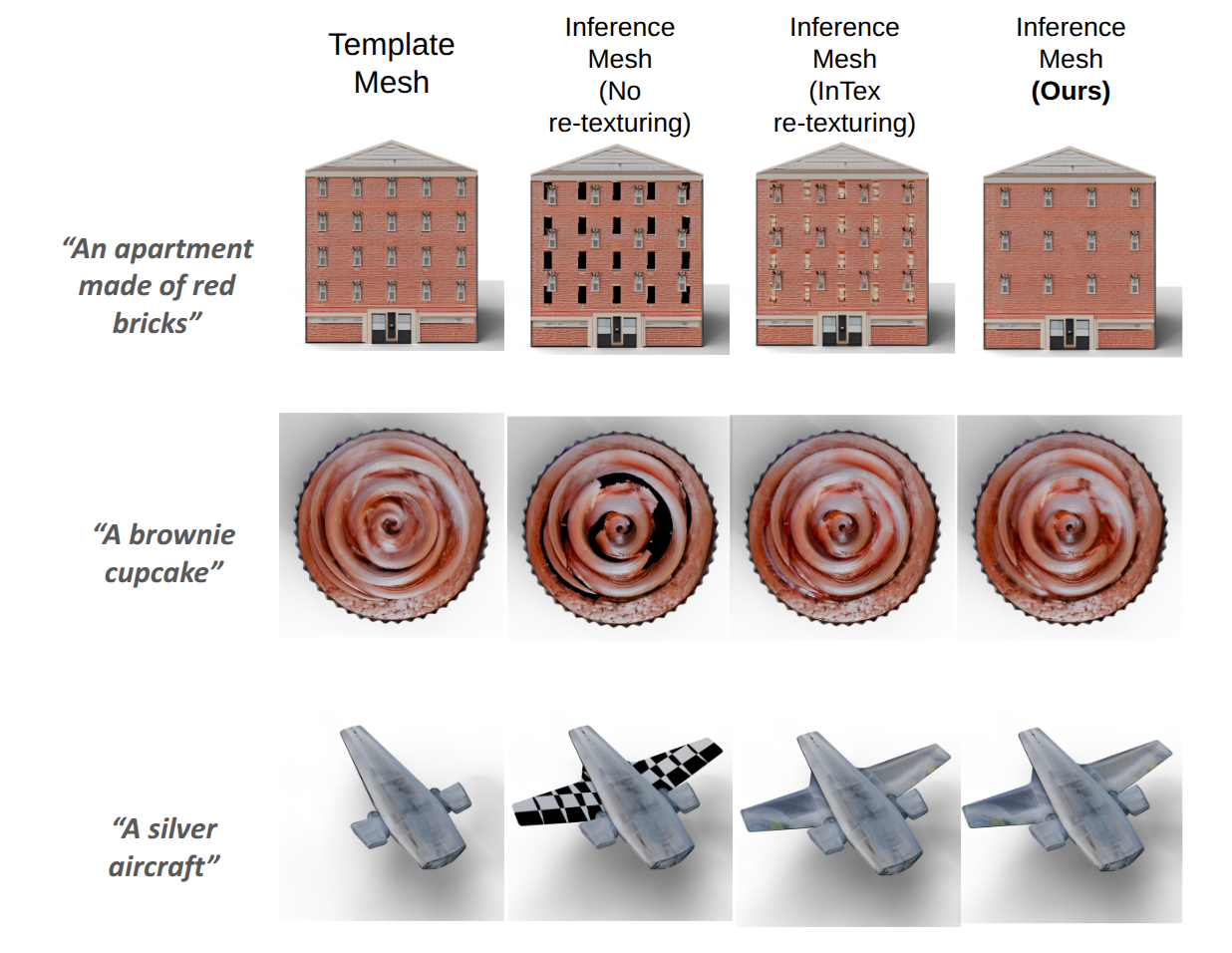}    
    \caption{Re-texturing results on dis-occluded regions, part removal and part addition. Directly applying template textures without re-texturing leaves untextured or distorted regions, while InTex re-texturing introduces semantic artifacts such as misplaced windows. Our method preserves both structure and local semantics across procedural variations.}
    \label{fig:part_texture}
\end{figure}

Although ProcTex can call the depth-aware inpainting network from \cite{tang2024intex}, we find that the inpainting will still drift in style, as depicted in Fig. \ref{fig:part_texture}. To ensure strict visual consistency, ProcTex falls back to exemplar-based transfer using PatchMatch \cite{barnes2009patchmatch}. For every view, ProcTex restricts patch candidates to pixels projected from the same component, and then slightly dilates the inpainting mask with a fixed sphere kernel to overwrite baked lighting artifacts and fill small edge gaps. This procedure robustly synthesizes coherent textures for the dis-occluded regions.

\subsection{ProcCorrNet for real-time texture transfer} \label{uv_displacement}
With the procedures in Sec. \ref{texture_generation} and \ref{texture_transfer}, ProcTex already supports text-guided texture synthesis over an entire procedural shape family. However, solving dense correspondence via cage deformation still takes up to a few minutes per component group, which blocks users to explore the parameter space at interactive rates. Existing learning-based methods handle only pair-wise alignments and do not generalize to a full parametric family of procedural shapes. We therefore introduce ProcCorrNet, a correspondence prediction network designed specifically for procedural modeling. To the best of our knowledge, ProcCorrNet is the first network that learns to predict dense correspondences between procedurally generated geometries and a designated template, enabling real-time texture transfer without iterative optimization. 

One challenge of estimating dense correspondence for procedural models is that the procedural shapes may exhibit large degree of geometric deformation, and even minor parameter changes can alter topology or vertex layout. Training a single model across all components is unstable, so we train a separate ProcCorrNet $NN_j$ for each component group $G_j$. This per-group design leverages the bounded variation within each component group while avoiding collapse across dissimilar structures (e.g., sofa legs vs. armrests). We describe the architecture and training pipeline for ProcCorrNet below.

Each ProcCorrNet adopts a simple yet effective PointNet-style architecture. The network takes as input a set of 3D point coordinates 
$(x,y,z)$ sampled from the source component surface. Firstly, a 4-layer MLP with 128 hidden units and ReLU activations is used to extract 256-dimensional local point features. A max-pooling operation is then used across all points to produce a 256-dimensional global descriptor, which is concatenated with each point’s local feature to form a 512-dimensional joint representation. This representation is then processed by a second 4-layer MLP with 256 hidden units and ReLU activation to predict per-point correspondences on the template surface. Importantly, the network’s outputs are also 3D coordinates in $(x,y,z)$ space, which are supervised to lie close to the surface of the template mesh and effectively resemble dense correspondence points.

During preprocessing, we generate high-quality but expensive correspondences using cage-based deformation. Each training pair consists of sampled surface points $P_i$ from a procedural component and their cage-mapped targets $P_{i,\text{transformed}}$ on the template. The network $f_\theta$ takes $P_i$ as input and outputs predicted correspondences $Y_i = f_\theta(P_i)$. 

We sample $5,000$ points per component per training iteration. The network is optimized with an L2 regression loss against $P_{i,\text{transformed}}$.

\begin{equation}
\mathcal{L} = \frac{1}{n_S} \sum_{k=1}^{n_S} | Y_i(k) - P_{i,\text{transformed}}(k) |_2^2
\end{equation}

At inference, given an unseen procedural mesh, ProcTex ProcTex segments it into component groups following the steps in Sec. \ref{part_identification_grouping}. Then, ProcTex applies the corresponding ProcCorrNet $NN_j$ to predict dense correspondences in a single feedforward pass. These predictions are projected to the nearest template surface points and converted to UV coordinates by barycentric interpolation. Importantly, this eliminates the need for optimization, making the process well-suited for real-time applications. 

ProcCorrNet is the core module in ProcTex for achieving interactive text-to-texture synthesis for procedural modeling. It amortizes the cost of expensive optimization into a one-time offline training phase, generalizes across the parameter space, and provides robust correspondences that are sufficiently accurate for texture transfer. While predictions are not perfectly aligned at every vertex, minor discrepancies are naturally mitigated by nearest-surface projection and texture lookup. In our experiments, ProcCorrNet reduces correspondence computation from several minutes per component with cage-based optimization to less than one second for all components combined, including the time for nearest-surface lookup, with only negligible loss in texture quality.

Beyond ProcTex, ProcCorrNet also provides a general framework for fast correspondence learning in procedural modeling. While other downstream tasks such as deformation transfer or animation may require stricter accuracy than texture transfer, ProcCorrNet demonstrates that dense correspondence for procedural components can be predicted in real time, eliminating the need for costly optimization and opening the door to a wide range of interactive applications.
\section{Results and Evaluations}
    We conduct experiments on a selection of 22 procedural models available on Blender Market, public repositories, and Infinigen \cite{raistrick2023infinite, infinigen2024indoors}. These models represent a diverse range of object types and offer varying levels of control over geometry. More details about the procedural models are shown in Appendix \ref{section:appendixA}. The training details of ProcCorrNet can be found in Appendix \ref{section:appendixB}. We analyze texture generation quality and visual consistency in Section \ref{qualitative_analysis} and \ref{quantitative_analysis}, present qualitative evaluation of the modified cage-based deformation in Section \ref{cage_ablation}, quantitatively analyze the performance of ProCorrNet in Section \ref{proccorrnet}, present ProcTex's part-level editability in Section \ref{part-level-edit}, and analyze ProcTex's runtime performance in Section \ref{interactive-analysis}.
    
    \subsection{Qualitative Analysis of Texture transfer} 
    \label{qualitative_analysis}
    
\begin{figure*}[]
    \centering
    \includegraphics[width=\linewidth]{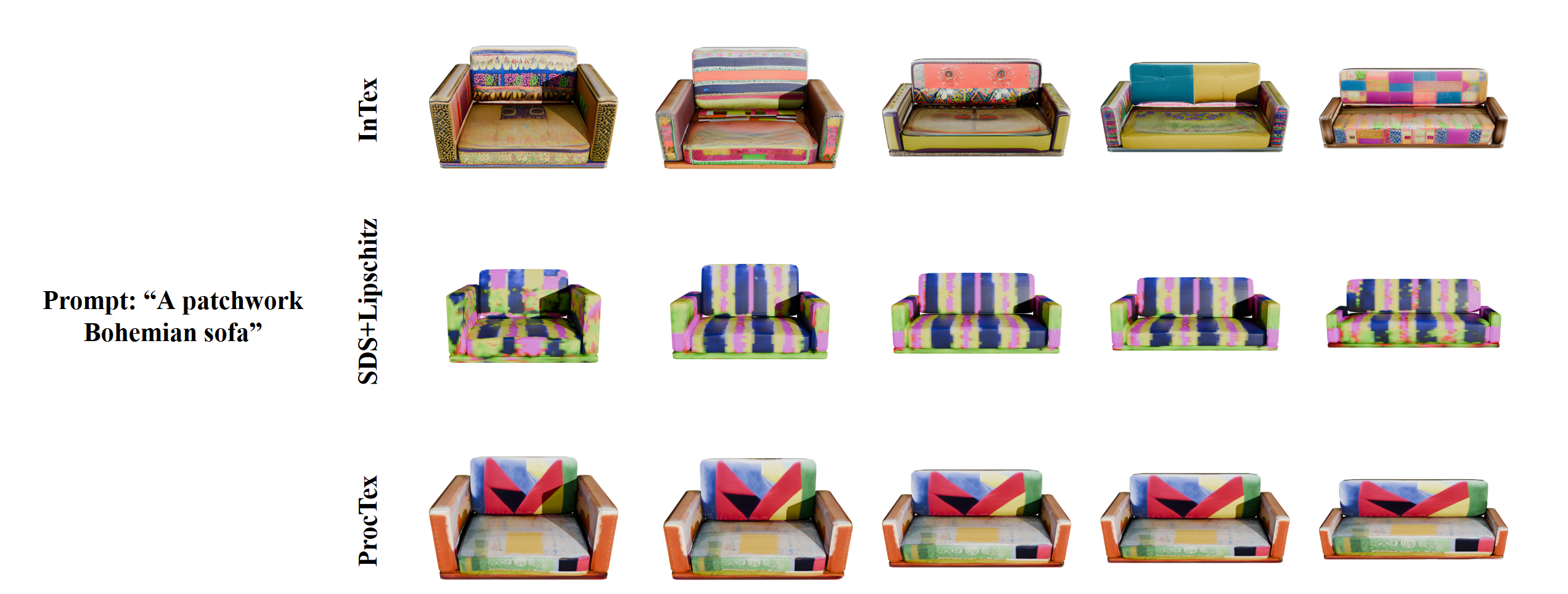}
    \vspace{0.7em}
    \rule{\linewidth}{0.4pt} 
    \vspace{0.7em}
    \includegraphics[width=\linewidth]{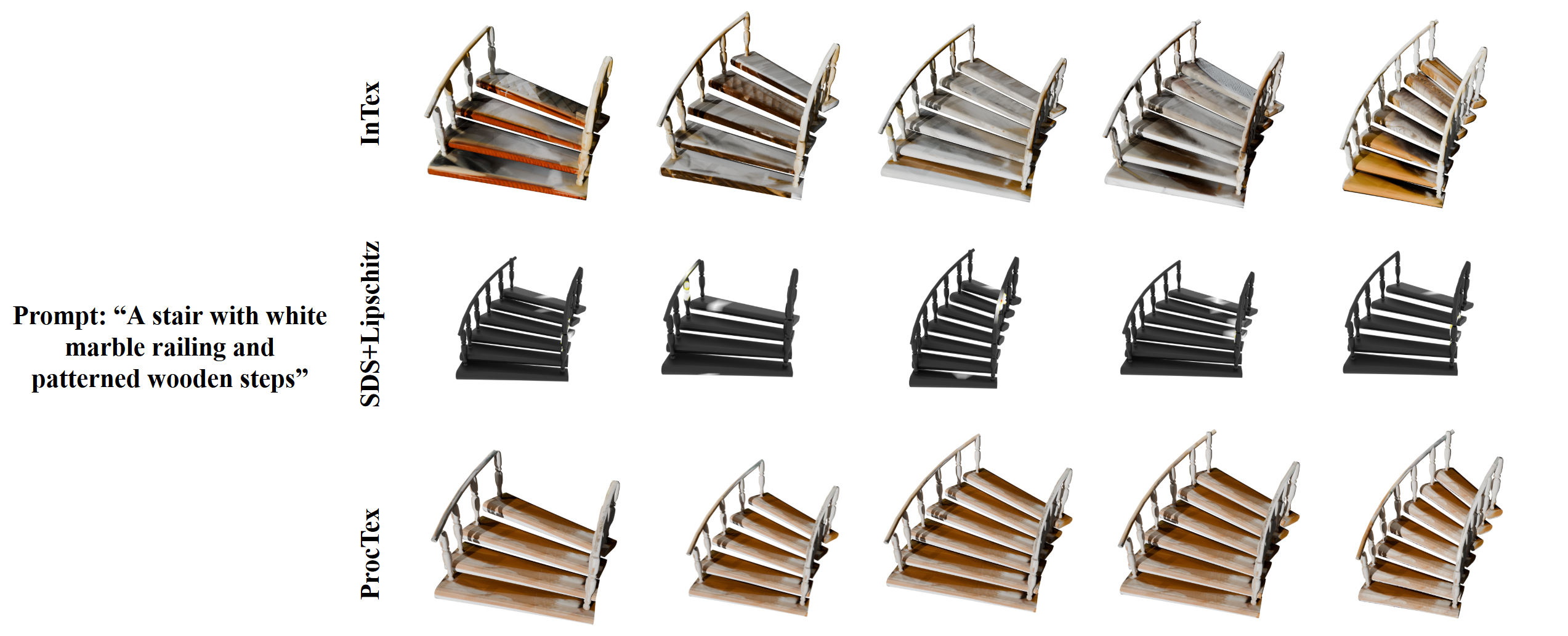}
    \rule{\linewidth}{0.4pt}
    \includegraphics[width=\linewidth]{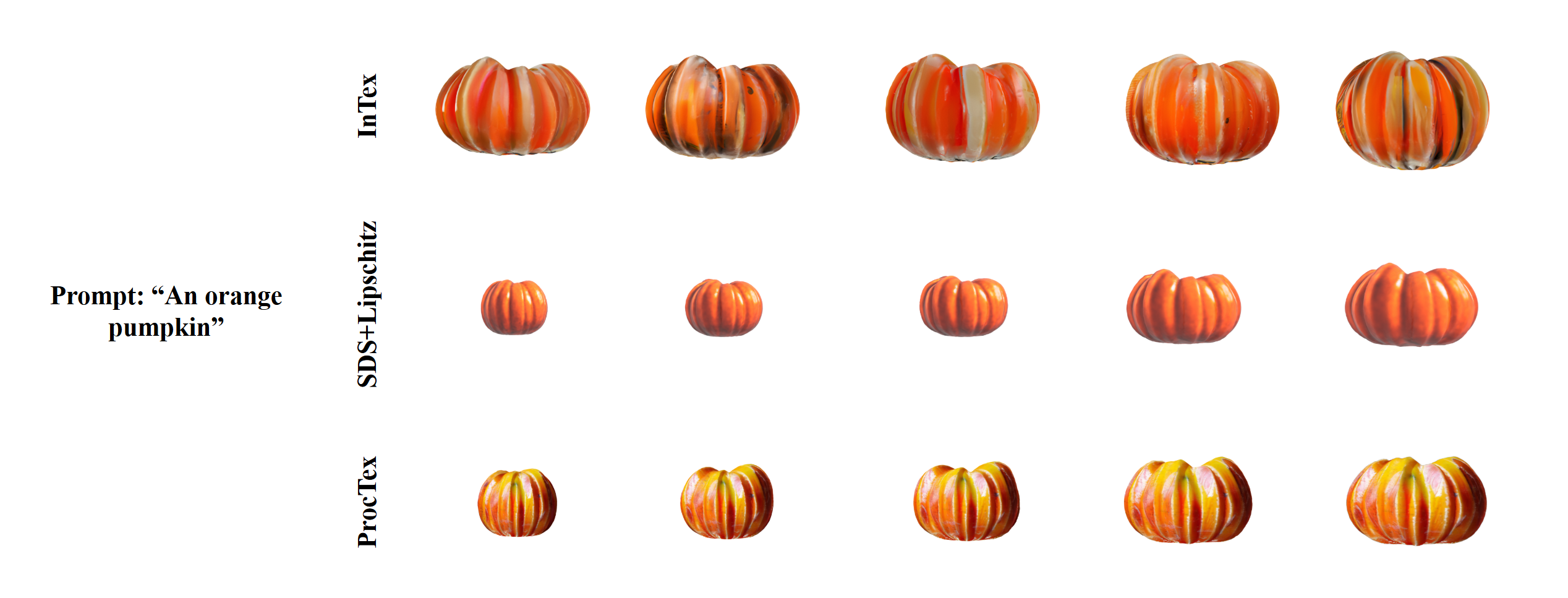}
    \caption{Qualitative comparison of ProcTex against two baselines (InTex and SDS+Lipschitz) across three prompts. ProcTex produces more consistent and detailed textures across procedural variations, while the baselines often yield distorted or incomplete results.}
    \label{fig:main_qualitative}
\end{figure*}

    We present textured outputs from ProcTex in Fig.~\ref{fig:teaser} and compare them with baseline methods in Fig.~\ref{fig:main_qualitative}. Additional examples are shown in Fig.~\ref{fig:nine_by_six_grid}. All results are generated using ProCorrNet correspondences at inference time. We refer readers to the supplemental video for the most faithful visual results, where we also provide real-time playbacks of ProcTex inference that demonstrates ProcTex's interactive capability.
    
    For the baselines, we first apply vanilla InTex to each procedural mesh independently. As there is no prior method tailored for text-guided texture synthesis across procedural shapes, we adapt a texture-field optimization approach as a secondary baseline. Specifically, we train an MLP following the design of \cite{Michel_2022_CVPR} to predict per-vertex colors using SDS~\cite{poole2022dreamfusion} supervision. The texture field is conditioned on procedural parameters and regularized with Lipschitz constraints~\cite{liu2022learningsmoothneuralfunctions} to encourage smoothness. To improve visual quality, the SDS loss is guided by depth-conditioned Stable Diffusion with ControlNet~\cite{rombach2022high, zhang2023adding}.

    We experimented with stronger SDS variants, including VSD~\cite{wang2024prolificdreamer} and SJC~\cite{wang2023score}, but found that these only increased training time substantially without noticeable improvement in appearance. The reported SDS+Lipschitz results thus reflect a carefully tuned pipeline aimed specifically at the goal of training a joint texture field for multiple procedural shapes. While this approach achieves some degree of cross-shape consistency, it often loses fine-scale texture fidelity. We hypothesize that this is an inherent limitation of optimizing a shared texture field across multiple varying geometries, as the optimization tends to even out high-frequency details, which leads to smoother but less detailed textures compared to ProcTex.

    As shown in Fig.~\ref{fig:main_qualitative}, InTex produces plausible textures on individual meshes but fails to preserve consistency across procedural variations as it needs to re-run texture generation for every single mesh. The SDS+Lipschitz baseline provides somewhat more consistent textures across variations but suffers from lower overall fidelity, limited detail, and frequent degenerate outputs as shown in the sofa example in Fig. ~\ref{fig:main_qualitative}. In particular, it struggles to converge on complex geometries undergoing non-affine transformations, as evident in the stair example. Moreover, this approach requires hours per prompt for training, and meshes must be subdivided to $\sim$100k vertices to avoid blurry outputs, leading to substantial memory overhead. Despite these resource demands, texture inconsistency still remains (e.g., sofa back patterns vary noticeably across shapes).

    In contrast, ProcTex achieves both high quality and strong consistency. By synthesizing a single texture map per component group, it enforces coherent appearance across procedural variations while maintaining detail within each mesh. The method produces stable, high-quality textures across all tested families. Together with the supplemental real-time demonstrations, these results highlight that ProcTex not only improves visual fidelity and efficiency but also enables interactive texture transfer, making it well-suited for iterative procedural modeling workflows.
    
    \begin{figure}[t]
    \centering
    \begin{subfigure}[t]{0.3\linewidth}
        \centering
        \includegraphics[width=\linewidth]{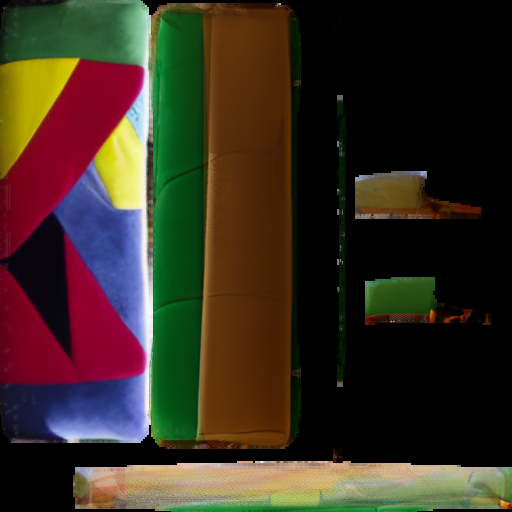}
    \end{subfigure}
    \hfill
    \begin{subfigure}[t]{0.30\linewidth}
        \centering
        \includegraphics[width=\linewidth]
        {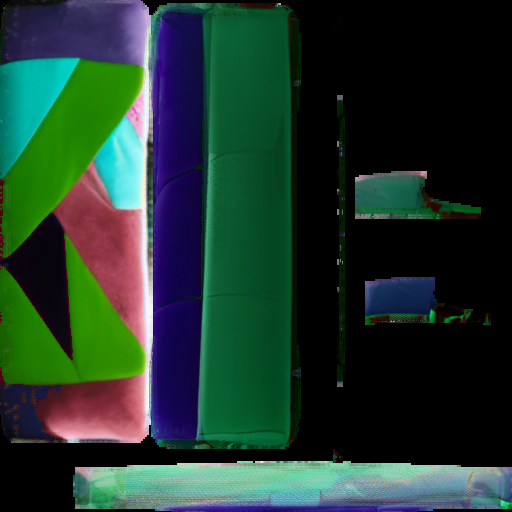}
    \end{subfigure}
    \hfill
    \begin{subfigure}[t]{0.30\linewidth}
        \centering
        \includegraphics[width=\linewidth]
        {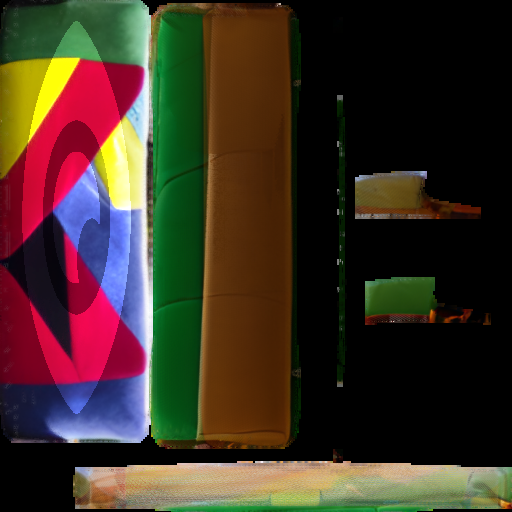}
    \end{subfigure}

    \vspace{1em} 
    \begin{subfigure}[t]{0.32\linewidth}
        \centering
        \includegraphics[width=\linewidth]
        {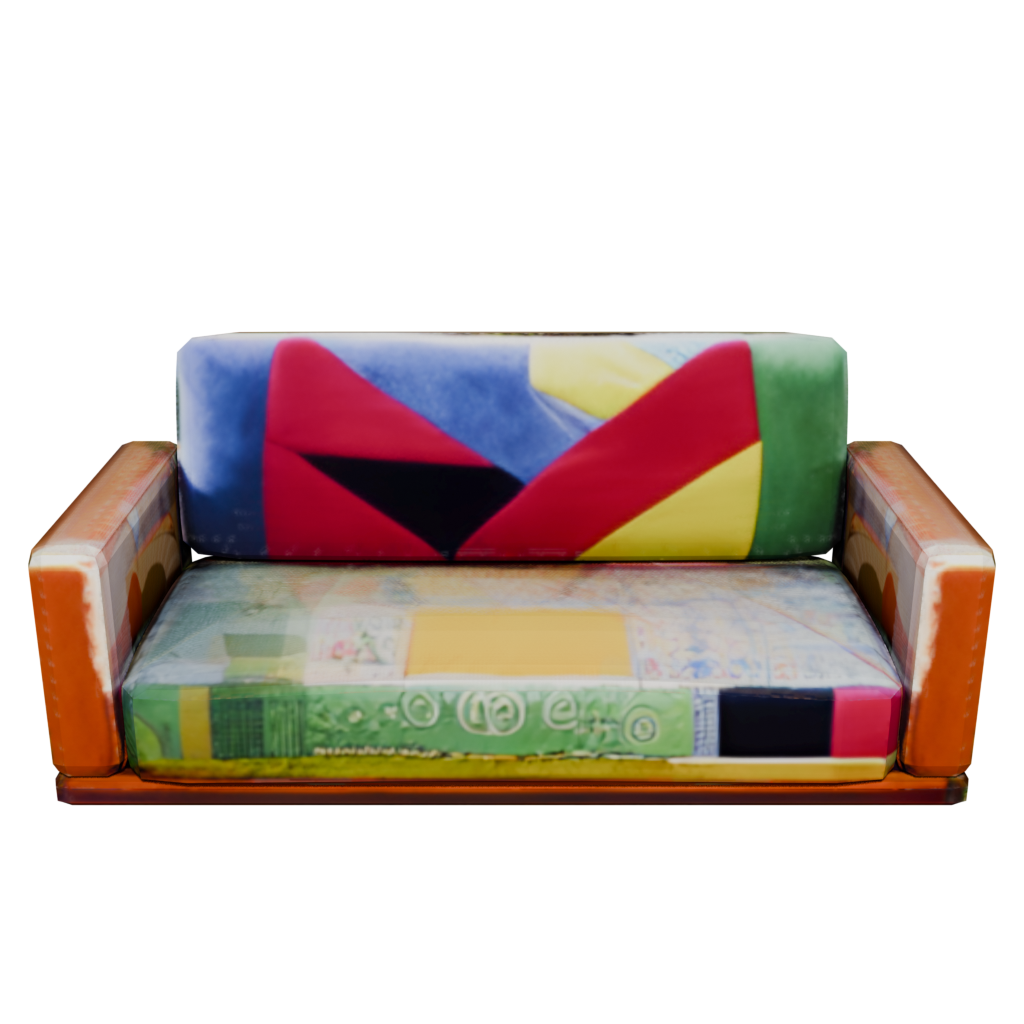}
    \end{subfigure}
    \hfill
    \begin{subfigure}[t]{0.32\linewidth}
        \centering
        \includegraphics[width=\linewidth]
        {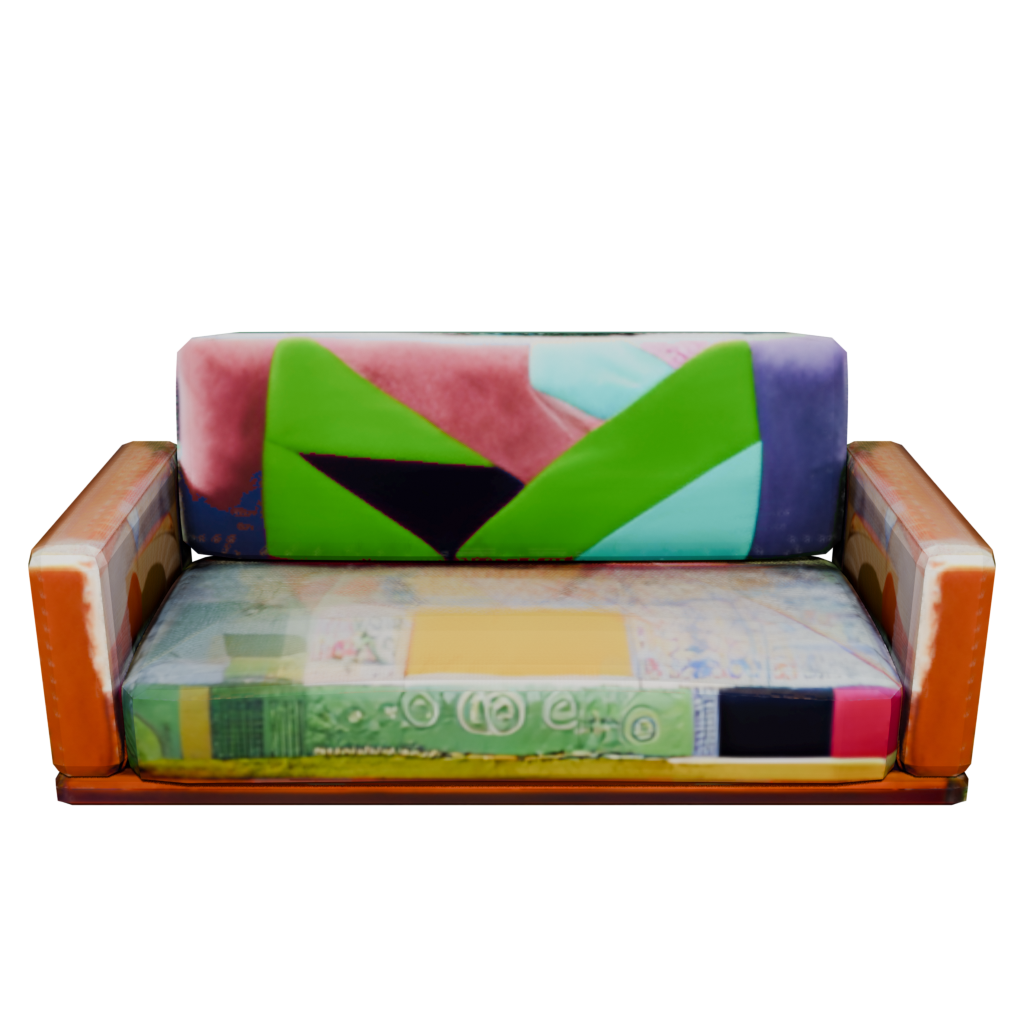}
    \end{subfigure}
    \hfill
    \begin{subfigure}[t]{0.32\linewidth}
        \centering
        \includegraphics[width=\linewidth]
        {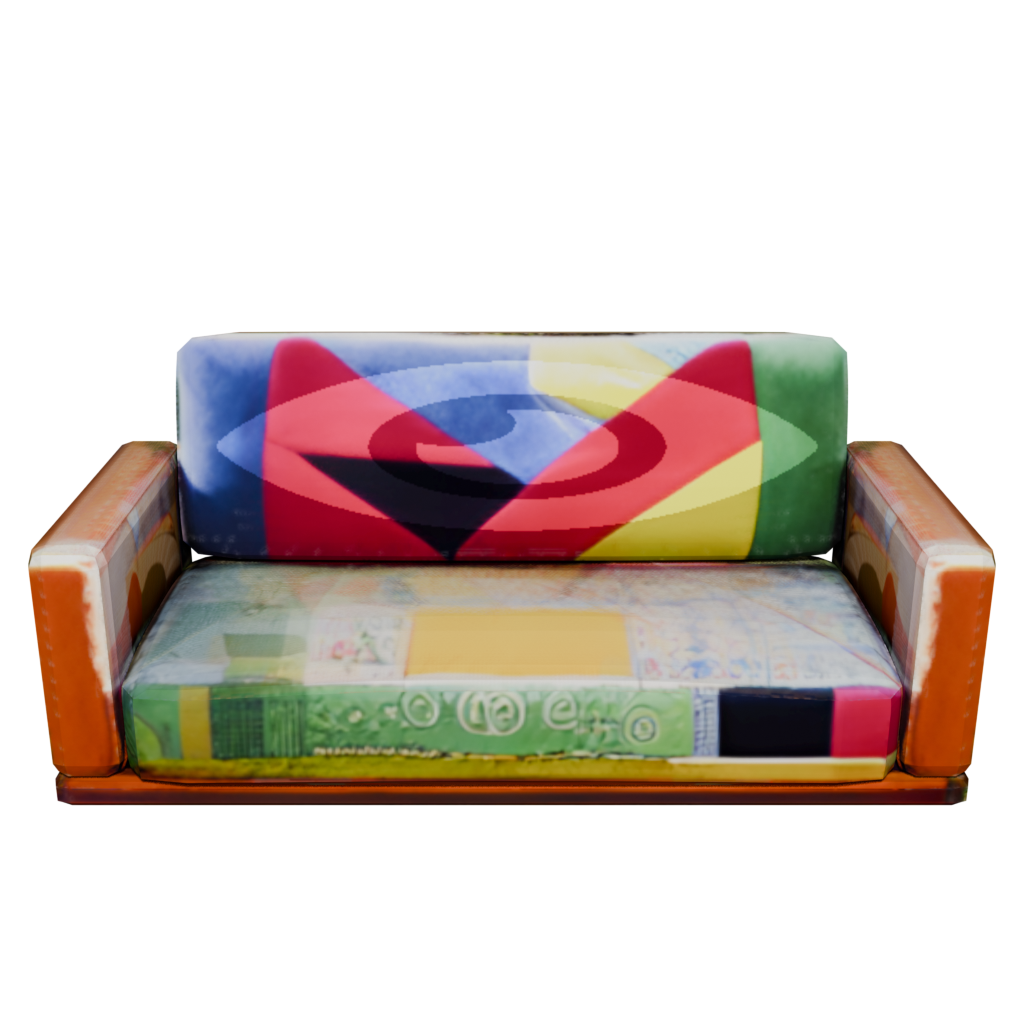}
    \end{subfigure}

    \caption{\textbf{Part appearance editing with texture maps generated by ProcTex.} Modify the sofa's back support texture. Left: Original texture; Middle: Texture after a hue shift of $+60^\circ$ in HSV color space; Right: Texture embossed with an eye-shaped logo}
    \label{fig:part_edit}
    \end{figure}

    \begin{table}[t!]
        \centering
        \renewcommand{\arraystretch}{1.4}
        \small
        \begin{tabular}{l ccc} \toprule
        \textbf{Model} & \textbf{LPIPS$\downarrow$} & \textbf{Tracked Point Texture Variance$\downarrow$}  \\ \hline
        InTex    & 0.467    & 0.439       \\ \hline
        SDS+Lipschitz       & 0.261     & 0.015     \\ \hline
        ProcTex        & \textbf{0.160}     & \textbf{0.006}        \\ \hline
        \bottomrule
        \end{tabular}
        \caption{Quantitative evaluation of texture transfer performance using perceptual (LPIPS) and correspondence-based (tracked point texture variance) metrics. Lower values indicate better visual similarity and texture consistency. ProcTex achieves the best performance across both measures, demonstrating its ability to preserve both global perceptual similarity and local texture consistency across procedural shape variations.}
        \label{table:quant_texture_transfer_table}
    \end{table}

    \begin{figure}[t]
    \centering
    \includegraphics[width=\linewidth,height=0.8\textheight,keepaspectratio]{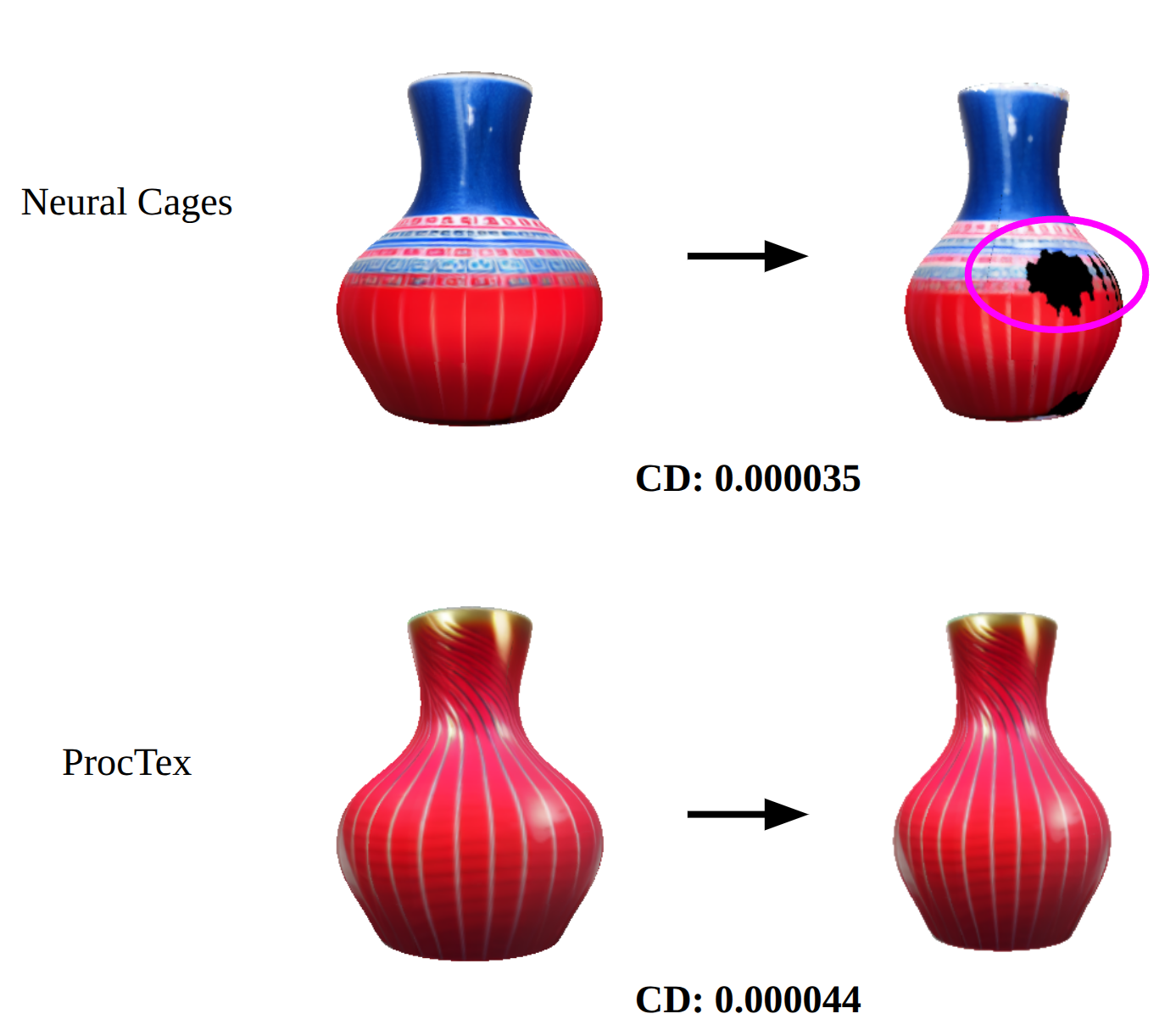}
    \caption{Qualitative comparison of Neural Cages and our modified variant on thin procedural models. The baseline Neural Cage often maps interior surfaces onto the exterior, which creates untextured regions (circled in purple). Our modification resolves these errors, ensuring complete texture coverage and higher visual fidelity, even though standard quantitative correspondence metrics like Chamfer distances (CD) remain nearly identical.}
    \label{fig:ablation_nerual_cage_qualitative}
    \end{figure}

    \subsection{Quantitative Analysis of Texture Transfer} \label{quantitative_analysis}
    To further evaluate the texture transfer performance, we introduce two quantitative metrics to measure the visual similarity and texture consistency across the procedural outputs. 
    
    For visual similarity, we compute the Learned Perceptual Image Patch Similarity (LPIPS) \cite{zhang2018perceptual} between textured renderings of the template mesh and its corresponding procedural meshes. Specifically, for each template mesh, we generate six renderings from canonical viewpoints (top, bottom, left, right, front, and back). For each procedural model, we randomly sample 20 meshes whose parameter configuration are not seen during training, apply ProcTex to produce their textured outputs, and render them under the same six viewpoints. We then compute LPIPS scores between each procedural mesh and its corresponding template mesh for the same canonical view. The final LPIPS score is obtained by averaging over all viewpoints, sampled meshes, and procedural models. 
    
    For texture consistency, we compute the color variance of tracked surface points of template meshes and their corresponding surface points on the procedural meshes. We begin by uniformly sampling 5,000 surface points on each template mesh and obtaining their corresponded points on procedural meshes using the modified cage-based deformation. For each tracked point on the template mesh and its corresponding points on other procedural meshes, we obtain their texture colors by barycentrically interpolating the UV-mapped texture. We then compute the standard deviation across the three RGB channels, and define the tracked point variance as the average of these values over all sampled points. The final score is reported as the mean across all procedural models.
    
    The quantitative results are shown in Table~\ref{table:quant_texture_transfer_table}. ProcTex consistently achieves the lowest LPIPS (0.160) and tracked point variance (0.006), indicating that it produces textures that are both perceptually closer to the template renderings and locally more consistent at corresponding surface points. Compared to InTex, which records significantly higher LPIPS (0.467) and variance (0.439), ProcTex demonstrates that naive inpainting-based texture filling struggles to preserve appearance across procedural variations. SDS+Lipschitz improves over InTex, particularly in tracked point variance (0.015), due to its smoothness regularization. However, the smoothness constraint alone does not guarantee that the optimized texture field will preserve textures across different shapes, let alone failure cases where input geometries undergo significant deformations such as stairs. As a result, its perceptual similarity (LPIPS 0.261) remains substantially higher than ProcTex.
    
    Overall, these results confirm that robust dense correspondence, together with our texture transfer strategy, is critical for preserving both global perceptual similarity and local texture consistency in the context of text-to-texture generation for procedural models.

    \subsection{Evaluation of the Modified Cage-based Deformation} \label{cage_ablation}
    We evaluate the effect of our modification to Neural Cages on procedural models with thin structures. In these cases, the baseline Neural Cage \cite{yifan2020neural} often maps interior surfaces onto the exterior, which produces large untextured regions on the visible surface (Fig.~\ref{fig:ablation_nerual_cage_qualitative}, top). Our modified variant resolves these artifacts by preserving the inside–outside distinction, leading to complete texture coverage and visually faithful results (Fig.~\ref{fig:ablation_nerual_cage_qualitative}, bottom). Both methods achieve very low Chamfer distances (0.000035 for the baseline and 0.000044 for the modified version), confirming strong geometric alignment in either case. However, only the modified variant secures reliable texture transfer, which is essential for high-quality visual results.

    \subsection{Quantitative Evaluation of ProCorrNet} \label{proccorrnet}
    \begin{table}[t!]
        \centering
        \renewcommand{\arraystretch}{1.2}
        \small
        \begin{tabular}{l ccc} \toprule
        \textbf{Model} & \textbf{Average point loss} & \textbf{Max point loss}  \\ \hline
        Bowl        & 0.014    & 0.040       \\ \hline
        Cake    & 0.017     & 0.026       \\ \hline
        Clam    & 0.016     & 0.018       \\ \hline
        Chair       & 0.009     & 0.032     \\ \hline
        Cupcake       & 0.018     & 0.022     \\ \hline
        Desk       & 0.009     & 0.020     \\ \hline
        Fork       & 0.006     & 0.010     \\ \hline
        Mugs       & 0.006     & 0.014     \\ \hline
        Mushroom       & 0.010     & 0.019     \\ \hline
        Mussel       & 0.004     & 0.016     \\ \hline
        Pumpkin & 0.014     & 0.071     \\ \hline
        Rock & 0.016     & 0.023     \\ \hline 
        Shell & 0.003     & 0.012     \\ \hline
        Sofa  & 0.011     & 0.025     \\ \hline
        Stair  & 0.014     & 0.018     \\ \hline
        Starfruit  & 0.008     & 0.023     \\ \hline
        Vase       &  0.006     & 0.014   \\ \hline
        
        \bottomrule
        \end{tabular}
        \caption{For each procedural model, we report the average point error and maximum point error across all sampled shapes. ProCorrNet consistently achieves low correspondence errors, indicating robust and accurate learning of dense mappings across diverse procedural families.}
        \label{table:proccorrnet_results}
    \end{table}

    We quantitatively evaluate the ability of ProCorrNet to learn dense correspondences for part-based procedural models. The set of procedural models used for testing is listed in Table \ref{table:proccorrnet_results}. For each procedural model, we construct ground-truth correspondences by sampling procedural variations and pairing each with the designated template shape through solving dense correspondence via modified cage-based deformation. A separate ProCorrNet is trained for each component group of the template. In cases where the number of components varies across shapes, we train networks according to the component structure of the template shape.

    We measure accuracy using the L2 correspondence error on procedural meshes that were unseen during training. Given a source point and its ground-truth mapped location on the template shape, we compute the error between this ground truth and the location predicted by ProCorrNet. We report both the average and largest errors across all points. The average point loss reflects overall accuracy across all vertices, while the max point loss captures the worst-case deviation. Avoiding such deviation is critical for texture transfer, as large local errors can cause visible texture distortions even if rare.
    
    The results are reported in Table \ref{table:proccorrnet_results}. Across all categories, ProCorrNet achieves consistently low average errors, all below 0.02 in the normalized xyz space, and bounded maximum deviations. Even in challenging families with large geometric variations, such as stairs and cakes, the predicted correspondences remain reliable. While ProCorrNet does not achieve perfect dense point-to-point alignment, we find the errors are generally tolerable and often negligible for texture transfer. This is because downstream steps such as nearest-surface projection, barycentric interpolation, and texture lookup inherently provide tolerance for slightly deviating correspondence points. Importantly, ProCorrNet provides dense correspondences at real-time inference speed, enabling interactive texture transfer with negligible impact on visual quality. These findings confirm that ProCorrNet produces reliable dense correspondences across a wide range of procedural models for the texture transfer task.
    
    \subsection{Part-level Editability} \label{part-level-edit}
    We further demonstrate ProcTex's capability for localized appearance editing in Fig. \ref{fig:part_edit}. Since ProcTex synthesizes individual texture maps for each component of the procedural model, edits such as color shifts or logo overlays can be applied to individual parts without unintended changes to other regions. This component-wise separation also produces a cleaner UV layout, which facilitates precise placement of edits.

    Beyond the examples shown, ProcTex accepts various forms of texture editing on the target texture, including regenerating content from entirely new text prompts. To update the results, one simply follows the procedure described in Sec. \ref{texture_baking} to rebake the source textures according to the modified target. The edited texture can then be applied consistently to meshes generated with any procedural parameter settings.

\newcommand{\sixpicsrow}[8]{%
  \begin{subfigure}[t]{0.155\textwidth}
    \centering\includegraphics[width=\linewidth]{#1/0.png}%
    \caption*{\scriptsize \parbox{0.9\linewidth}{\raggedright #2: #3}}
  \end{subfigure}%
  \begin{subfigure}[t]{0.155\textwidth}
    \centering\includegraphics[width=\linewidth]{#1/1.png}%
    \caption*{\scriptsize \parbox{0.9\linewidth}{\raggedright #2: #4}}
  \end{subfigure}%
  \begin{subfigure}[t]{0.155\textwidth}
    \centering\includegraphics[width=\linewidth]{#1/2.png}%
    \caption*{\scriptsize \parbox{0.9\linewidth}{\raggedright #2: #5}}
  \end{subfigure}%
  \begin{subfigure}[t]{0.155\textwidth}
    \centering\includegraphics[width=\linewidth]{#1/3.png}%
    \caption*{\scriptsize \parbox{0.9\linewidth}{\raggedright #2: #6}}
  \end{subfigure}%
  \begin{subfigure}[t]{0.155\textwidth}
    \centering\includegraphics[width=\linewidth]{#1/4.png}%
    \caption*{\scriptsize \parbox{0.9\linewidth}{\raggedright #2: #7}}
  \end{subfigure}%
  \begin{subfigure}[t]{0.155\textwidth}
    \centering\includegraphics[width=\linewidth]{#1/5.png}%
    \caption*{\scriptsize \parbox{0.9\linewidth}{\raggedright #2: #8}}
  \end{subfigure}%
  \\[0.3em]%
}

\begin{figure*}[]
  \centering
  \sixpicsrow{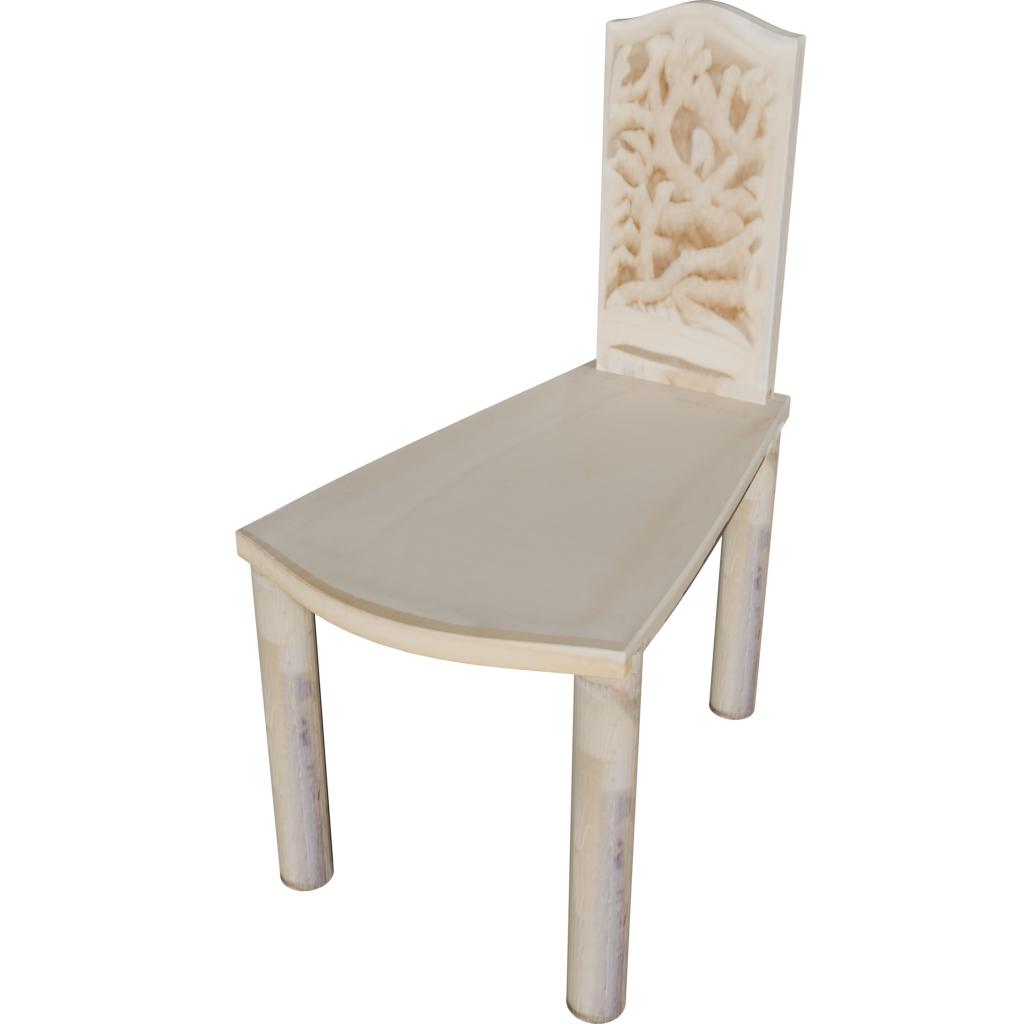}{Chair}
  {back\_height=0.3, width=0.3, ...} {back\_height=0.375, width=0.425, ...} {back\_height=0.442, width=0.425, ...}
  {back\_height=0.525, width=0.55, ...}
  {back\_height=0.525, width=0.675, ...}
  {back\_height=0.6, width=0.8, ...}
  \sixpicsrow{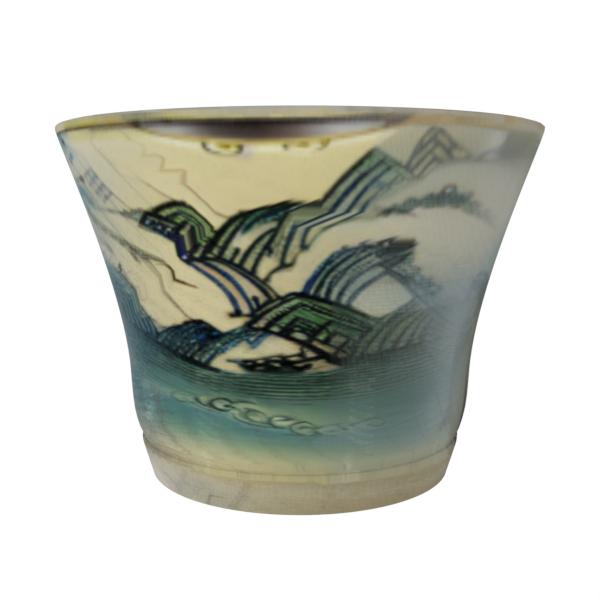}{Pottery}
  {base\_radius=95mm, height=1} 
  {base\_radius=90mm, height=1.5}  
  {base\_radius=85mm, height=2} 
  {base\_radius=80mm, height=2.5}  
  {base\_radius=75mm, height=3}  
  {base\_radius=70mm, height=3.5} 
  \sixpicsrow{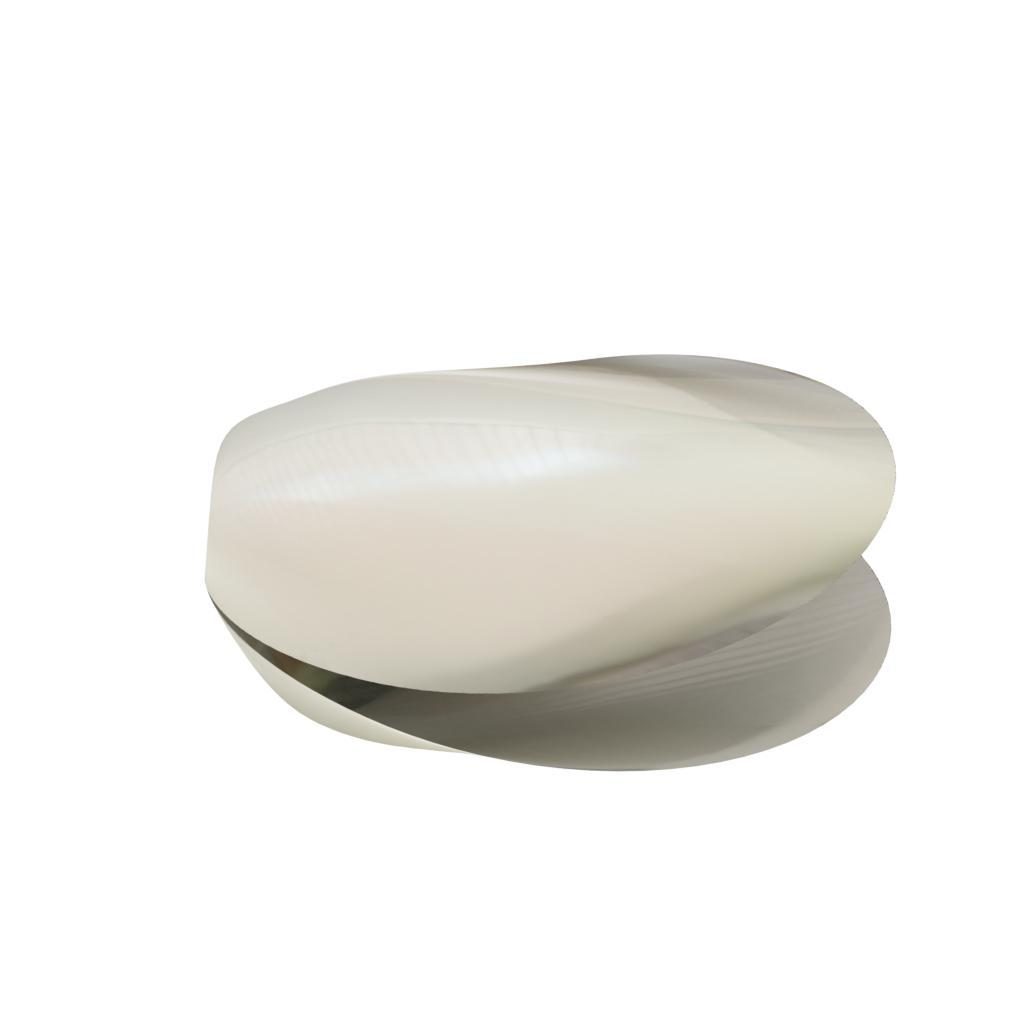}{Shell}
  {open\_angle=0, curvature=0.5}  
  {open\_angle=$\frac{\pi}{10}$, curvature=0.55} 
  {open\_angle=$\frac{\pi}{5}$, curvature=0.6} 
  {open\_angle=$\frac{3\pi}{10}$, curvature=0.65} 
  {open\_angle=$\frac{2\pi}{5}$, curvature=0.7} 
  {open\_angle=$\frac{\pi}{2}$, curvature=0.75} 
  \sixpicsrow{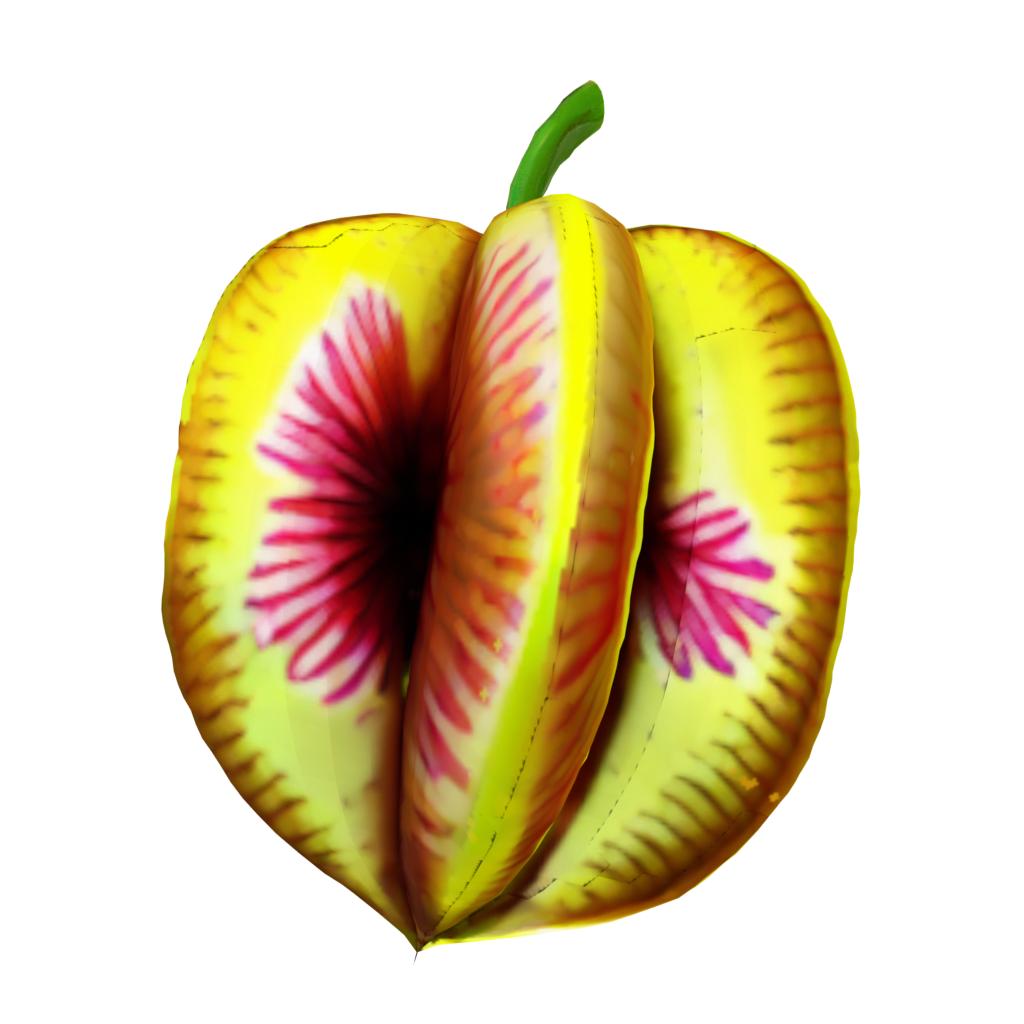}{Starfruit}
  {radius\_1 = 0.45, radius\_2 = 0.7} 
  {radius\_1 = 0.5, radius\_2 = 0.75} 
  {radius\_1 = 0.55, radius\_2 = 0.8} 
  {radius\_1 = 0.6, radius\_2 = 0.85} 
  {radius\_1 = 0.65, radius\_2 = 0.9} 
  {radius\_1 = 0.7, radius\_2 = 0.95} 
  \sixpicsrow{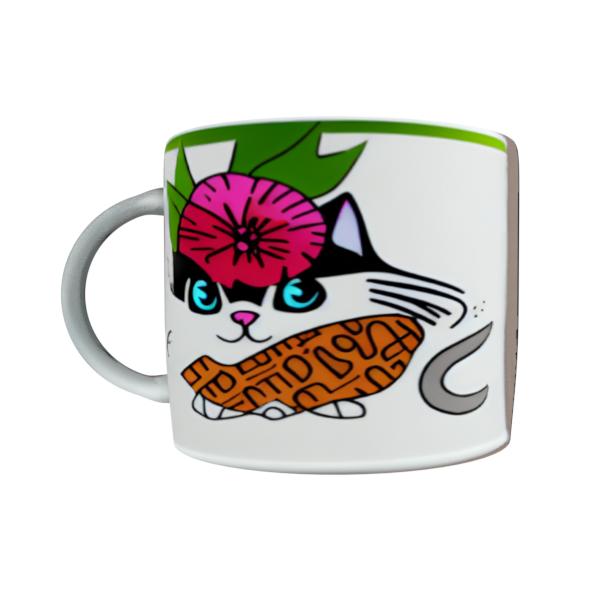}{Mug}
  {radius=2.0, height=3.00} 
  {radius=1.8, height=3.25}  
  {radius=1.6, height=3.50} 
  {radius=1.4, height=3.75}  
  {radius=1.2, height=4.00}  
  {radius=1.3, height=4.25} 
  \sixpicsrow{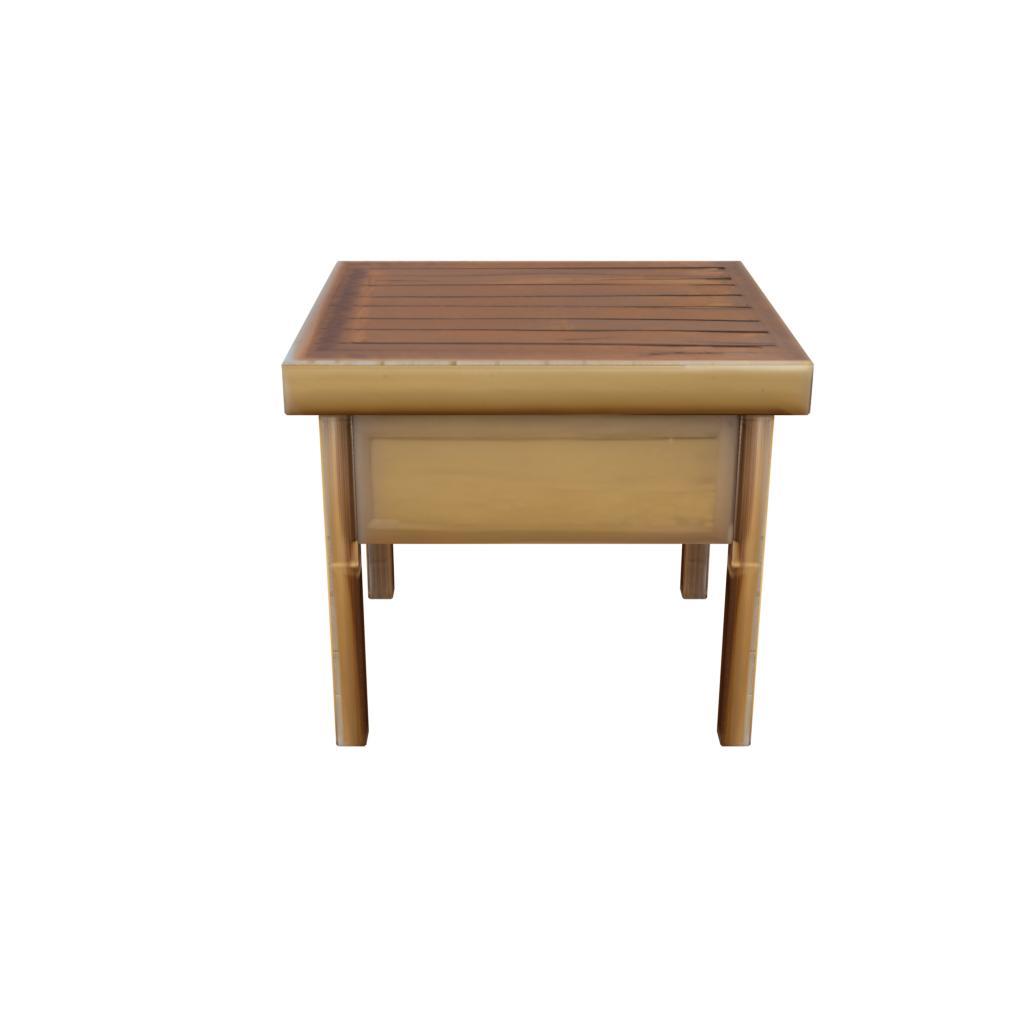}{Desk}
  {w=1, h=1, drawer\_size=0.325} 
  {w=1, h=1, drawer\_size=0.375}
  {w=1.25, h=1, drawer\_size=0.325}
  {w=1.4, h=1.5, drawer\_size=0.375}
  {w=1.4, h=1.5, drawer\_size=0.375}
  {w=1.8, h=2.5, drawer\_size=0.45}
  \caption{More qualitative results. Each row corresponds to one procedural model. Text prompts used from top to bottom: (1) “An ivory chair with brown patterns in the backrest”; (2) “A Japanese ceramic pottery”; (3) "A white shell"; (4) "A yellow starfruit"; (5) "A mug with cat patterns"; (6) "A worn school desk". For the chair model, only a subset of parameters are displayed here; please refer to the supplemental video for more details. These results highlight that our method maintains both global style and local detail across wide procedural variations.}
  \label{fig:nine_by_six_grid}
\end{figure*}

    \begin{table}[t!]
        \centering
        \renewcommand{\arraystretch}{1.2}
        \small
        \begin{tabular}{l ccc} \toprule
        \textbf{Model} & \textbf{Training Time} & \textbf{Inference Time}  \\ \hline
        InTex    & 10-12s     & 10-12s       \\ \hline
        SDS+Lipschitz       & 2-5hours     & 0.1s     \\ \hline
        ProcTex        & 10-50min     & 0.5-1s        \\ \hline
        \bottomrule
        \end{tabular}
        \caption{Training and inference runtimes of ProcTex and baseline methods. InTex requires per-shape re-synthesis (10–12s), while SDS+Lipschitz achieves fast inference but unreliable in quality. ProcTex achieves both consistent textures and interactive inference (0.5–1s), enabling real-time procedural modeling workflows.}
        \label{table:training_inference_speed}
    \end{table}

    \subsection{Runtime Performance and Interactivity Anlysis} 
    \label{interactive-analysis}
    Table \ref{table:training_inference_speed} summarizes the training and inference costs of ProcTex compared with baseline methods. InTex requires re-synthesizing textures for every new mesh, leading to inference times on the order of 10–12 seconds per shape, which is too slow for interactive design. The SDS+Lipschitz baseline achieves fast inference (0.1s) by directly optimizing a global texture field, but its results are inconsistent across procedural variations and often fail to converge on complex geometries, making it impractical for real use. ProcTex strikes a balance between speed and quality. It requires a moderate one-time training cost of 10–50 minutes, after which textures can be transferred in 0.5–1 second per shape. This enables both real-time feedback and consistent, high-quality texture synthesis across procedural models. Such interactive performance is essential in procedural modeling workflows, where users explore large design spaces and need immediate visual feedback to guide parameter adjustments.

\section{Conclusions}

We propose ProcTex, a system that effectively enables text-to-texture synthesis for part-based procedural models. Our model is capable of generating high-quality textures with remarkable visual consistency across a diverse set of procedural families of shapes. We also introduce ProcCorrNet, which effectively learns dense surface correspondence for procedural models to perform real-time texture transfer at inference. Moreover, our approach enables local appearance editing through part-level texture synthesis, and also includes a robust re-texturing pipeline that supports structural geometric variations in procedural modeling.

\bibliographystyle{eg-alpha-doi} 
\bibliography{egbibsample}       

\newcommand{\etalchar}[1]{$^{#1}$}
\begin{thebibliography}{\uppercase{MZWVG07}}

\bibitem[ASOW23]{abdelreheem2023SATR}
\textsc{Abdelreheem A., Skorokhodov I., Ovsjanikov M., Wonka P.}:
\newblock Satr: Zero-shot semantic segmentation of 3d shapes.
\newblock In \emph{Proceedings of the International Conference on Computer Vision ({ICCV})} (2023).

\bibitem[AVE22]{albergo2022building}
\textsc{Albergo M.~S., Vanden-Eijnden E.}:
\newblock Building normalizing flows with stochastic interpolants.
\newblock \emph{arXiv preprint arXiv:2209.15571} (2022).

\bibitem[BKA{\etalchar{*}}24]{bensadoun2024meta3dtexturegenfast}
\textsc{Bensadoun R., Kleiman Y., Azuri I., Harosh O., Vedaldi A., Neverova N., Gafni O.}:
\newblock Meta 3d texturegen: Fast and consistent texture generation for 3d objects, 2024.
\newblock URL: \url{https://arxiv.org/abs/2407.02430}, \href {http://arxiv.org/abs/2407.02430} {\path{arXiv:2407.02430}}.

\bibitem[BM92]{ICP}
\textsc{Besl P., McKay N.~D.}:
\newblock A method for registration of 3-d shapes.
\newblock \emph{IEEE Transactions on Pattern Analysis and Machine Intelligence 14}, 2 (1992), 239--256.
\newblock \href {https://doi.org/10.1109/34.121791} {\path{doi:10.1109/34.121791}}.

\bibitem[BSFG09]{barnes2009patchmatch}
\textsc{Barnes C., Shechtman E., Finkelstein A., Goldman D.~B.}:
\newblock Patchmatch: A randomized correspondence algorithm for structural image editing.
\newblock \emph{ACM Trans. Graph. 28}, 3 (2009), 24.

\bibitem[CCJJ23]{chen2023fantasia3d}
\textsc{Chen R., Chen Y., Jiao N., Jia K.}:
\newblock Fantasia3d: Disentangling geometry and appearance for high-quality text-to-3d content creation.
\newblock In \emph{Proceedings of the IEEE/CVF International Conference on Computer Vision (ICCV)} (October 2023), pp.~22246--22256.

\bibitem[CKF{\etalchar{*}}23]{cao2023texfusion}
\textsc{Cao T., Kreis K., Fidler S., Sharp N., Yin K.}:
\newblock Texfusion: Synthesizing 3d textures with text-guided image diffusion models.
\newblock In \emph{Proceedings of the IEEE/CVF International Conference on Computer Vision (ICCV)} (2023).

\bibitem[Com18]{blender}
\textsc{Community B.~O.}:
\newblock \emph{Blender - a 3D modelling and rendering package}.
\newblock Blender Foundation, Stichting Blender Foundation, Amsterdam, 2018.
\newblock URL: \url{http://www.blender.org}.

\bibitem[CRB23]{cao2023unsupervised}
\textsc{Cao D., Roetzer P., Bernard F.}:
\newblock Unsupervised learning of robust spectral shape matching.
\newblock \emph{ACM Transactions on Graphics (TOG)} (2023).
\newblock URL: \url{https://doi.org/10.1145/3592107}, \href {https://doi.org/10.1145/3592107} {\path{doi:10.1145/3592107}}.

\bibitem[CSL{\etalchar{*}}23]{chen2023text2tex}
\textsc{Chen D.~Z., Siddiqui Y., Lee H.-Y., Tulyakov S., Nie{\ss}ner M.}:
\newblock Text2tex: Text-driven texture synthesis via diffusion models.
\newblock In \emph{Proceedings of the IEEE/CVF International Conference on Computer Vision} (2023), pp.~18558--18568.

\bibitem[DKD{\etalchar{*}}22]{deng2022unsupervised}
\textsc{Deng B., Kulal S., Dong Z., Deng C., Tian Y., Wu J.}:
\newblock Unsupervised learning of shape programs with repeatable implicit parts.
\newblock \emph{Advances in Neural Information Processing Systems 35} (2022), 37837--37850.

\bibitem[DLAH23]{decatur2023paintbrush}
\textsc{Decatur D., Lang I., Aberman K., Hanocka R.}:
\newblock 3d paintbrush: Local stylization of 3d shapes with cascaded score distillation.
\newblock \emph{arXiv} (2023).

\bibitem[DOW{\etalchar{*}}24]{deng2024flashtex}
\textsc{Deng K., Omernick T., Weiss A., Ramanan D., Zhu J.-Y., Zhou T., Agrawala M.}:
\newblock Flashtex: Fast relightable mesh texturing with lightcontrolnet.
\newblock In \emph{European Conference on Computer Vision (ECCV)} (2024).

\bibitem[DYM{\etalchar{*}}24]{dong2024coin3d}
\textsc{Dong W., Yang B., Ma L., Liu X., Cui L., Bao H., Ma Y., Cui Z.}:
\newblock Coin3d: Controllable and interactive 3d assets generation with proxy-guided conditioning, 2024.
\newblock \href {http://arxiv.org/abs/2405.08054} {\path{arXiv:2405.08054}}.

\bibitem[ELC19]{eisenberger2019smoothshellsmultiscaleshape}
\textsc{Eisenberger M., Lähner Z., Cremers D.}:
\newblock Smooth shells: Multi-scale shape registration with functional maps, 2019.
\newblock URL: \url{https://arxiv.org/abs/1905.12512}, \href {http://arxiv.org/abs/1905.12512} {\path{arXiv:1905.12512}}.

\bibitem[ENP{\etalchar{*}}19]{ellis2019write}
\textsc{Ellis K., Nye M., Pu Y., Sosa F., Tenenbaum J., Solar-Lezama A.}:
\newblock Write, execute, assess: Program synthesis with a repl.
\newblock \emph{Advances in Neural Information Processing Systems 32} (2019).

\bibitem[Gha08]{ghali2008constructive}
\textsc{Ghali S.}:
\newblock Constructive solid geometry.
\newblock \emph{Introduction to geometric computing} (2008), 277--283.

\bibitem[GHS{\etalchar{*}}22]{guerrero2022matformer}
\textsc{Guerrero P., Ha{\v{s}}an M., Sunkavalli K., M{\v{e}}ch R., Boubekeur T., Mitra N.~J.}:
\newblock Matformer: A generative model for procedural materials.
\newblock \emph{arXiv preprint arXiv:2207.01044} (2022).

\bibitem[GHX{\etalchar{*}}24]{ganeshan2024parsel}
\textsc{Ganeshan A., Huang R., Xu X., Jones R.~K., Ritchie D.}:
\newblock Parsel: Parameterized shape editing with language.
\newblock \emph{ACM Transactions on Graphics (TOG) 43}, 6 (2024), 1--14.

\bibitem[GJL{\etalchar{*}}24]{gao2024genesistexadaptingimagedenoising}
\textsc{Gao C., Jiang B., Li X., Zhang Y., Yu Q.}:
\newblock Genesistex: Adapting image denoising diffusion to texture space, 2024.
\newblock URL: \url{https://arxiv.org/abs/2403.17782}, \href {http://arxiv.org/abs/2403.17782} {\path{arXiv:2403.17782}}.

\bibitem[GPAM{\etalchar{*}}14]{goodfellow2014generative}
\textsc{Goodfellow I.~J., Pouget-Abadie J., Mirza M., Xu B., Warde-Farley D., Ozair S., Courville A., Bengio Y.}:
\newblock Generative adversarial nets.
\newblock \emph{Advances in neural information processing systems 27} (2014).

\bibitem[HGZ{\etalchar{*}}24]{huo2024texgen}
\textsc{Huo D., Guo Z., Zuo X., Shi Z., Lu J., Dai P., Xu S., Cheng L., Yang Y.-H.}:
\newblock Texgen: Text-guided 3d texture generation with multi-view sampling and resampling.
\newblock In \emph{European Conference on Computer Vision} (2024), Springer, pp.~352--368.

\bibitem[HZG{\etalchar{*}}24]{hong2024lrmlargereconstructionmodel}
\textsc{Hong Y., Zhang K., Gu J., Bi S., Zhou Y., Liu D., Liu F., Sunkavalli K., Bui T., Tan H.}:
\newblock Lrm: Large reconstruction model for single image to 3d, 2024.
\newblock URL: \url{https://arxiv.org/abs/2311.04400}, \href {http://arxiv.org/abs/2311.04400} {\path{arXiv:2311.04400}}.

\bibitem[JBX{\etalchar{*}}20]{jones2020shapeassembly}
\textsc{Jones R.~K., Barton T., Xu X., Wang K., Jiang E., Guerrero P., Mitra N.~J., Ritchie D.}:
\newblock Shapeassembly: Learning to generate programs for 3d shape structure synthesis.
\newblock \emph{ACM Transactions on Graphics (TOG) 39}, 6 (2020), 1--20.

\bibitem[JSW05]{mvc}
\textsc{Ju T., Schaefer S., Warren J.}:
\newblock Mean value coordinates for closed triangular meshes.
\newblock In \emph{ACM SIGGRAPH 2005 Papers} (New York, NY, USA, 2005), SIGGRAPH '05, Association for Computing Machinery, p.~561–566.
\newblock URL: \url{https://doi.org/10.1145/1186822.1073229}, \href {https://doi.org/10.1145/1186822.1073229} {\path{doi:10.1145/1186822.1073229}}.

\bibitem[Kel21]{kelly2021cityengine}
\textsc{Kelly T.}:
\newblock Cityengine: An introduction to rule-based modeling.
\newblock \emph{Urban informatics} (2021), 637--662.

\bibitem[KKLD23]{kerbl3Dgaussians}
\textsc{Kerbl B., Kopanas G., Leimk{\"u}hler T., Drettakis G.}:
\newblock 3d gaussian splatting for real-time radiance field rendering.
\newblock \emph{ACM Transactions on Graphics 42}, 4 (July 2023).
\newblock URL: \url{https://repo-sam.inria.fr/fungraph/3d-gaussian-splatting/}.

\bibitem[KLA{\etalchar{*}}24]{kim2024meshup}
\textsc{Kim H., Lang I., Aigerman N., Groueix T., Kim V.~G., Hanocka R.}:
\newblock Meshup: Multi-target mesh deformation via blended score distillation.
\newblock \emph{arXiv preprint arXiv:2408.14899} (2024).

\bibitem[KXBT22]{khalid2022clipmesh}
\textsc{Khalid N.~M., Xie T., Belilovsky E., Tiberiu P.}:
\newblock Clip-mesh: Generating textured meshes from text using pretrained image-text models.
\newblock \emph{SIGGRAPH Asia 2022 Conference Papers} (December 2022).

\bibitem[LCBH{\etalchar{*}}22]{lipman2022flow}
\textsc{Lipman Y., Chen R.~T., Ben-Hamu H., Nickel M., Le M.}:
\newblock Flow matching for generative modeling.
\newblock \emph{arXiv preprint arXiv:2210.02747} (2022).

\bibitem[LGT{\etalchar{*}}23]{lin2023magic3d}
\textsc{Lin C.-H., Gao J., Tang L., Takikawa T., Zeng X., Huang X., Kreis K., Fidler S., Liu M.-Y., Lin T.-Y.}:
\newblock Magic3d: High-resolution text-to-3d content creation.
\newblock In \emph{Proceedings of the IEEE/CVF Conference on Computer Vision and Pattern Recognition} (2023), pp.~300--309.

\bibitem[LLL{\etalchar{*}}24]{liu2024part123}
\textsc{Liu A., Lin C., Liu Y., Long X., Dou Z., Guo H.-X., Luo P., Wang W.}:
\newblock Part123: Part-aware 3d reconstruction from a single-view image.
\newblock In \emph{ACM SIGGRAPH 2024 Conference Papers} (2024), pp.~1--12.

\bibitem[LRR{\etalchar{*}}17]{litany2017deepfunctionalmapsstructured}
\textsc{Litany O., Remez T., Rodolà E., Bronstein A.~M., Bronstein M.~M.}:
\newblock Deep functional maps: Structured prediction for dense shape correspondence, 2017.
\newblock URL: \url{https://arxiv.org/abs/1704.08686}, \href {http://arxiv.org/abs/1704.08686} {\path{arXiv:1704.08686}}.

\bibitem[LWH{\etalchar{*}}23]{liu2023zero1to3}
\textsc{Liu R., Wu R., Hoorick B.~V., Tokmakov P., Zakharov S., Vondrick C.}:
\newblock Zero-1-to-3: Zero-shot one image to 3d object, 2023.
\newblock \href {http://arxiv.org/abs/2303.11328} {\path{arXiv:2303.11328}}.

\bibitem[LWJ{\etalchar{*}}22]{liu2022learningsmoothneuralfunctions}
\textsc{Liu H.-T.~D., Williams F., Jacobson A., Fidler S., Litany O.}:
\newblock Learning smooth neural functions via lipschitz regularization, 2022.
\newblock URL: \url{https://arxiv.org/abs/2202.08345}, \href {http://arxiv.org/abs/2202.08345} {\path{arXiv:2202.08345}}.

\bibitem[LXLW24]{liu2024text}
\textsc{Liu Y., Xie M., Liu H., Wong T.-T.}:
\newblock Text-guided texturing by synchronized multi-view diffusion.
\newblock In \emph{SIGGRAPH Asia 2024 Conference Papers} (2024), pp.~1--11.

\bibitem[LXZ{\etalchar{*}}23]{lorraine2023att3d}
\textsc{Lorraine J., Xie K., Zeng X., Lin C.-H., Takikawa T., Sharp N., Lin T.-Y., Liu M.-Y., Fidler S., Lucas J.}:
\newblock Att3d: Amortized text-to-3d object synthesis.
\newblock In \emph{Proceedings of the IEEE/CVF International Conference on Computer Vision} (2023), pp.~17946--17956.

\bibitem[LYC{\etalchar{*}}25]{lin2025kiss3dgen}
\textsc{Lin J., Yang X., Chen M., Xu Y., Yan D., Wu L., Xu X., Xu L., Zhang S., Chen Y.-C.}:
\newblock Kiss3dgen: Repurposing image diffusion models for 3d asset generation.
\newblock \emph{arXiv preprint arXiv:2503.01370} (2025).

\bibitem[LZL{\etalchar{*}}25]{li2025triposg}
\textsc{Li Y., Zou Z.-X., Liu Z., Wang D., Liang Y., Yu Z., Liu X., Guo Y.-C., Liang D., Ouyang W., et~al.}:
\newblock Triposg: High-fidelity 3d shape synthesis using large-scale rectified flow models.
\newblock \emph{arXiv preprint arXiv:2502.06608} (2025).

\bibitem[MBOL{\etalchar{*}}21]{text2mesh}
\textsc{Michel O., Bar-On R., Liu R., Benaim S., Hanocka R.}:
\newblock Text2mesh: Text-driven neural stylization for meshes.
\newblock \emph{arXiv preprint arXiv:2112.03221} (2021).

\bibitem[MBOL{\etalchar{*}}22]{Michel_2022_CVPR}
\textsc{Michel O., Bar-On R., Liu R., Benaim S., Hanocka R.}:
\newblock Text2mesh: Text-driven neural stylization for meshes.
\newblock In \emph{Proceedings of the IEEE/CVF Conference on Computer Vision and Pattern Recognition (CVPR)} (June 2022), pp.~13492--13502.

\bibitem[MRP{\etalchar{*}}23]{metzer2023latent}
\textsc{Metzer G., Richardson E., Patashnik O., Giryes R., Cohen-Or D.}:
\newblock Latent-nerf for shape-guided generation of 3d shapes and textures.
\newblock In \emph{Proceedings of the IEEE/CVF conference on computer vision and pattern recognition} (2023), pp.~12663--12673.

\bibitem[MSK10]{merrell2010computer}
\textsc{Merrell P., Schkufza E., Koltun V.}:
\newblock Computer-generated residential building layouts.
\newblock In \emph{ACM SIGGRApH Asia 2010 papers}. 2010, pp.~1--12.

\bibitem[MST{\etalchar{*}}21]{mildenhall2021nerf}
\textsc{Mildenhall B., Srinivasan P.~P., Tancik M., Barron J.~T., Ramamoorthi R., Ng R.}:
\newblock Nerf: Representing scenes as neural radiance fields for view synthesis.
\newblock \emph{Communications of the ACM 65}, 1 (2021), 99--106.

\bibitem[MVG13]{martinovic2013bayesian}
\textsc{Martinovic A., Van~Gool L.}:
\newblock Bayesian grammar learning for inverse procedural modeling.
\newblock In \emph{Proceedings of the IEEE Conference on Computer Vision and Pattern Recognition} (2013), pp.~201--208.

\bibitem[MWH{\etalchar{*}}06]{muller2006procedural}
\textsc{M{\"u}ller P., Wonka P., Haegler S., Ulmer A., Van~Gool L.}:
\newblock Procedural modeling of buildings.
\newblock In \emph{ACM SIGGRAPH 2006 Papers}. 2006, pp.~614--623.

\bibitem[MZWVG07]{muller2007image}
\textsc{M{\"u}ller P., Zeng G., Wonka P., Van~Gool L.}:
\newblock Image-based procedural modeling of facades.
\newblock \emph{ACM Trans. Graph. 26}, 3 (2007), 85.

\bibitem[NBA18]{nishida2018procedural}
\textsc{Nishida G., Bousseau A., Aliaga D.~G.}:
\newblock Procedural modeling of a building from a single image.
\newblock In \emph{Computer Graphics Forum} (2018), vol.~37, Wiley Online Library, pp.~415--429.

\bibitem[NGDA{\etalchar{*}}16]{nishida2016interactive}
\textsc{Nishida G., Garcia-Dorado I., Aliaga D.~G., Benes B., Bousseau A.}:
\newblock Interactive sketching of urban procedural models.
\newblock \emph{ACM Transactions on Graphics (TOG) 35}, 4 (2016), 1--11.

\bibitem[Per85]{perlin1985image}
\textsc{Perlin K.}:
\newblock An image synthesizer.
\newblock \emph{ACM Siggraph Computer Graphics 19}, 3 (1985), 287--296.

\bibitem[PHHM96]{prusinkiewicz1996systems}
\textsc{Prusinkiewicz P., Hammel M., Hanan J., Mech R.}:
\newblock L-systems: from the theory to visual models of plants.
\newblock In \emph{Proceedings of the 2nd CSIRO Symposium on Computational Challenges in Life Sciences} (1996), vol.~3, Citeseer, pp.~1--32.

\bibitem[PHL{\etalchar{*}}09]{palubicki2009self}
\textsc{Palubicki W., Horel K., Longay S., Runions A., Lane B., M{\v{e}}ch R., Prusinkiewicz P.}:
\newblock Self-organizing tree models for image synthesis.
\newblock \emph{ACM Transactions On Graphics (TOG) 28}, 3 (2009), 1--10.

\bibitem[PJBM22]{poole2022dreamfusion}
\textsc{Poole B., Jain A., Barron J.~T., Mildenhall B.}:
\newblock Dreamfusion: Text-to-3d using 2d diffusion.
\newblock \emph{arXiv preprint arXiv:2209.14988} (2022).

\bibitem[PLH88]{prusinkiewicz1988development}
\textsc{Prusinkiewicz P., Lindenmayer A., Hanan J.}:
\newblock Development models of herbaceous plants for computer imagery purposes.
\newblock In \emph{Proceedings of the 15th annual conference on Computer graphics and interactive techniques} (1988), pp.~141--150.

\bibitem[PLH{\etalchar{*}}25]{pearl2025geocode}
\textsc{Pearl O., Lang I., Hu Y., Yeh R.~A., Hanocka R.}:
\newblock Geocode: Interpretable shape programs.
\newblock In \emph{Computer Graphics Forum} (2025), vol.~44, Wiley Online Library, p.~e15276.

\bibitem[Pru86]{prusinkiewicz1986graphical}
\textsc{Prusinkiewicz P.}:
\newblock Graphical applications of l-systems.
\newblock In \emph{Proceedings of graphics interface} (1986), vol.~86, pp.~247--253.

\bibitem[QCG{\etalchar{*}}24]{qiu2024richdreamer}
\textsc{Qiu L., Chen G., Gu X., Zuo Q., Xu M., Wu Y., Yuan W., Dong Z., Bo L., Han X.}:
\newblock Richdreamer: A generalizable normal-depth diffusion model for detail richness in text-to-3d.
\newblock In \emph{Proceedings of the IEEE/CVF Conference on Computer Vision and Pattern Recognition} (2024), pp.~9914--9925.

\bibitem[QMH{\etalchar{*}}23]{qian2023magic123}
\textsc{Qian G., Mai J., Hamdi A., Ren J., Siarohin A., Li B., Lee H.-Y., Skorokhodov I., Wonka P., Tulyakov S., et~al.}:
\newblock Magic123: One image to high-quality 3d object generation using both 2d and 3d diffusion priors.
\newblock \emph{arXiv preprint arXiv:2306.17843} (2023).

\bibitem[RBL{\etalchar{*}}22]{rombach2022high}
\textsc{Rombach R., Blattmann A., Lorenz D., Esser P., Ommer B.}:
\newblock High-resolution image synthesis with latent diffusion models.
\newblock In \emph{Proceedings of the IEEE/CVF conference on computer vision and pattern recognition} (2022), pp.~10684--10695.

\bibitem[RKH{\etalchar{*}}21]{clip}
\textsc{Radford A., Kim J.~W., Hallacy C., Ramesh A., Goh G., Agarwal S., Sastry G., Askell A., Mishkin P., Clark J., Krueger G., Sutskever I.}:
\newblock Learning transferable visual models from natural language supervision.
\newblock \emph{CoRR abs/2103.00020} (2021).
\newblock URL: \url{https://arxiv.org/abs/2103.00020}, \href {http://arxiv.org/abs/2103.00020} {\path{arXiv:2103.00020}}.

\bibitem[RLM{\etalchar{*}}23]{raistrick2023infinite}
\textsc{Raistrick A., Lipson L., Ma Z., Mei L., Wang M., Zuo Y., Kayan K., Wen H., Han B., Wang Y., et~al.}:
\newblock Infinite photorealistic worlds using procedural generation.
\newblock In \emph{Proceedings of the IEEE/CVF conference on computer vision and pattern recognition} (2023), pp.~12630--12641.

\bibitem[RMA{\etalchar{*}}23]{richardson2023texture}
\textsc{Richardson E., Metzer G., Alaluf Y., Giryes R., Cohen-Or D.}:
\newblock Texture: Text-guided texturing of 3d shapes.
\newblock In \emph{ACM SIGGRAPH 2023 conference proceedings} (2023), pp.~1--11.

\bibitem[RMK{\etalchar{*}}24]{infinigen2024indoors}
\textsc{Raistrick A., Mei L., Kayan K., Yan D., Zuo Y., Han B., Wen H., Parakh M., Alexandropoulos S., Lipson L., Ma Z., Deng J.}:
\newblock Infinigen indoors: Photorealistic indoor scenes using procedural generation.
\newblock In \emph{Proceedings of the IEEE/CVF Conference on Computer Vision and Pattern Recognition (CVPR)} (June 2024), pp.~21783--21794.

\bibitem[SCS{\etalchar{*}}22]{saharia2022photorealistic}
\textsc{Saharia C., Chan W., Saxena S., Li L., Whang J., Denton E.~L., Ghasemipour K., Gontijo~Lopes R., Karagol~Ayan B., Salimans T., et~al.}:
\newblock Photorealistic text-to-image diffusion models with deep language understanding.
\newblock \emph{Advances in neural information processing systems 35} (2022), 36479--36494.

\bibitem[SO20]{sharma2020weaklysuperviseddeepfunctional}
\textsc{Sharma A., Ovsjanikov M.}:
\newblock Weakly supervised deep functional map for shape matching, 2020.
\newblock URL: \url{https://arxiv.org/abs/2009.13339}, \href {http://arxiv.org/abs/2009.13339} {\path{arXiv:2009.13339}}.

\bibitem[Str06]{stroud2006boundary}
\textsc{Stroud I.}:
\newblock \emph{Boundary representation modelling techniques}.
\newblock Springer Science \& Business Media, 2006.

\bibitem[Sut64]{sutherland1964sketch}
\textsc{Sutherland I.~E.}:
\newblock Sketch pad a man-machine graphical communication system.
\newblock In \emph{Proceedings of the SHARE design automation workshop} (1964), pp.~6--329.

\bibitem[SWY{\etalchar{*}}23]{shi2023MVDream}
\textsc{Shi Y., Wang P., Ye J., Mai L., Li K., Yang X.}:
\newblock Mvdream: Multi-view diffusion for 3d generation.
\newblock \emph{arXiv:2308.16512} (2023).

\bibitem[Tea25]{hunyuan3d2025hunyuan3d}
\textsc{Team T.~H.}:
\newblock Hunyuan3d 2.1: From images to high-fidelity 3d assets with production-ready pbr material, 2025.
\newblock \href {http://arxiv.org/abs/2506.15442} {\path{arXiv:2506.15442}}.

\bibitem[TLC{\etalchar{*}}24]{tang2024intex}
\textsc{Tang J., Lu R., Chen X., Wen X., Zeng G., Liu Z.}:
\newblock Intex: Interactive text-to-texture synthesis via unified depth-aware inpainting.
\newblock \emph{arXiv preprint arXiv:2403.11878} (2024).

\bibitem[TRZ{\etalchar{*}}23]{tang2023dreamgaussian}
\textsc{Tang J., Ren J., Zhou H., Liu Z., Zeng G.}:
\newblock Dreamgaussian: Generative gaussian splatting for efficient 3d content creation.
\newblock \emph{arXiv preprint arXiv:2309.16653} (2023).

\bibitem[TZF{\etalchar{*}}24]{tang2024segmentmeshzeroshotmesh}
\textsc{Tang G., Zhao W., Ford L., Benhaim D., Zhang P.}:
\newblock Segment any mesh: Zero-shot mesh part segmentation via lifting segment anything 2 to 3d, 2024.
\newblock URL: \url{https://arxiv.org/abs/2408.13679}, \href {http://arxiv.org/abs/2408.13679} {\path{arXiv:2408.13679}}.

\bibitem[Ume91]{umeyamaicp}
\textsc{Umeyama S.}:
\newblock Least-squares estimation of transformation parameters between two point patterns.
\newblock \emph{IEEE Transactions on Pattern Analysis and Machine Intelligence 13}, 4 (1991), 376--380.
\newblock \href {https://doi.org/10.1109/34.88573} {\path{doi:10.1109/34.88573}}.

\bibitem[WDL{\etalchar{*}}23]{wang2023score}
\textsc{Wang H., Du X., Li J., Yeh R.~A., Shakhnarovich G.}:
\newblock Score jacobian chaining: Lifting pretrained 2d diffusion models for 3d generation.
\newblock In \emph{Proceedings of the IEEE/CVF Conference on Computer Vision and Pattern Recognition} (2023), pp.~12619--12629.

\bibitem[WLW{\etalchar{*}}24]{wang2024prolificdreamer}
\textsc{Wang Z., Lu C., Wang Y., Bao F., Li C., Su H., Zhu J.}:
\newblock Prolificdreamer: High-fidelity and diverse text-to-3d generation with variational score distillation.
\newblock \emph{Advances in Neural Information Processing Systems 36} (2024).

\bibitem[Wor96]{worley1996cellular}
\textsc{Worley S.}:
\newblock A cellular texture basis function.
\newblock In \emph{Proceedings of the 23rd annual conference on Computer graphics and interactive techniques} (1996), pp.~291--294.

\bibitem[WWF{\etalchar{*}}23]{wei2023taps3d}
\textsc{Wei J., Wang H., Feng J., Lin G., Yap K.-H.}:
\newblock Taps3d: Text-guided 3d textured shape generation from pseudo supervision.
\newblock In \emph{Proceedings of the IEEE/CVF conference on computer vision and pattern recognition} (2023), pp.~16805--16815.

\bibitem[WXZ21]{wu2021deepcad}
\textsc{Wu R., Xiao C., Zheng C.}:
\newblock Deepcad: A deep generative network for computer-aided design models.
\newblock In \emph{Proceedings of the IEEE/CVF International Conference on Computer Vision} (2021), pp.~6772--6782.

\bibitem[XLX{\etalchar{*}}24]{xiang2024structured}
\textsc{Xiang J., Lv Z., Xu S., Deng Y., Wang R., Zhang B., Chen D., Tong X., Yang J.}:
\newblock Structured 3d latents for scalable and versatile 3d generation.
\newblock \emph{arXiv preprint arXiv:2412.01506} (2024).

\bibitem[XPC{\etalchar{*}}21]{xu2021inferring}
\textsc{Xu X., Peng W., Cheng C.-Y., Willis K.~D., Ritchie D.}:
\newblock Inferring cad modeling sequences using zone graphs.
\newblock In \emph{Proceedings of the IEEE/CVF conference on computer vision and pattern recognition} (2021), pp.~6062--6070.

\bibitem[YAK{\etalchar{*}}20]{yifan2020neural}
\textsc{Yifan W., Aigerman N., Kim V.~G., Chaudhuri S., Sorkine-Hornung O.}:
\newblock Neural cages for detail-preserving 3d deformations.
\newblock In \emph{Proceedings of the IEEE/CVF conference on computer vision and pattern recognition} (2020), pp.~75--83.

\bibitem[YFW{\etalchar{*}}24]{yi2023gaussiandreamer}
\textsc{Yi T., Fang J., Wang J., Wu G., Xie L., Zhang X., Liu W., Tian Q., Wang X.}:
\newblock Gaussiandreamer: Fast generation from text to 3d gaussians by bridging 2d and 3d diffusion models.
\newblock In \emph{CVPR} (2024).

\bibitem[YHK{\etalchar{*}}24]{yeh2024texturedreamer}
\textsc{Yeh Y.-Y., Huang J.-B., Kim C., Xiao L., Nguyen-Phuoc T., Khan N., Zhang C., Chandraker M., Marshall C.~S., Dong Z., et~al.}:
\newblock Texturedreamer: Image-guided texture synthesis through geometry-aware diffusion.
\newblock \emph{arXiv preprint arXiv:2401.09416} (2024).

\bibitem[YYG{\etalchar{*}}24]{yu2024texgen}
\textsc{Yu X., Yuan Z., Guo Y.-C., Liu Y.-T., Liu J., Li Y., Cao Y.-P., Liang D., Qi X.}:
\newblock Texgen: a generative diffusion model for mesh textures.
\newblock \emph{ACM Trans. Graph. 43}, 6 (2024).
\newblock \href {https://doi.org/10.1145/3687909} {\path{doi:10.1145/3687909}}.

\bibitem[ZGL{\etalchar{*}}23]{zhou2023partslipenhancinglowshot3d}
\textsc{Zhou Y., Gu J., Li X., Liu M., Fang Y., Su H.}:
\newblock Partslip++: Enhancing low-shot 3d part segmentation via multi-view instance segmentation and maximum likelihood estimation, 2023.
\newblock URL: \url{https://arxiv.org/abs/2312.03015}, \href {http://arxiv.org/abs/2312.03015} {\path{arXiv:2312.03015}}.

\bibitem[ZIE{\etalchar{*}}18]{zhang2018perceptual}
\textsc{Zhang R., Isola P., Efros A.~A., Shechtman E., Wang O.}:
\newblock The unreasonable effectiveness of deep features as a perceptual metric.
\newblock In \emph{CVPR} (2018).

\bibitem[ZLZ{\etalchar{*}}24]{zhang2024scene}
\textsc{Zhang Y., Li Z., Zhou M., Wu S., Wu J.}:
\newblock The scene language: Representing scenes with programs, words, and embeddings.
\newblock \emph{arXiv preprint arXiv:2410.16770} (2024).

\bibitem[ZRA23]{zhang2023adding}
\textsc{Zhang L., Rao A., Agrawala M.}:
\newblock Adding conditional control to text-to-image diffusion models.
\newblock In \emph{Proceedings of the IEEE/CVF International Conference on Computer Vision} (2023), pp.~3836--3847.

\end{thebibliography}


\clearpage
\appendix

\section{Procedural Generators Statistics from Our Experiments} \label{section:appendixA}
We summarize the statistics of the procedural generators used in our experiments in Table. \ref{table:Generator Statistics}. For each generator, we report the object category, source, number of parameters covered by the training data, number of groups, whether re-texturing is required, and whether the topology varies.

\begin{table}[t!]
    \centering
    \small
    \renewcommand{\arraystretch}{1.4}
    \begin{tabular}{l ccccc} \toprule
    \textbf{Category} & \textbf{Source} & \textbf{\# params} & \textbf{\# Groups} & \textbf{Retex.?} & \textbf{Topology}  \\ \hline
    Pottery    & Blender     &  2 & 1 & No & Fixed     \\ 
    \hline
    Vase$_{0}$ & Blender     &  5 & 1 & No & Varies     \\ 
    \hline
    Vase$_{1}$ & Blender     &  2 & 1 & No & Fixed     \\ 
    \hline
    Mug        & Blender     &  1 & 2 & No  & Fixed     \\ 
    \hline
    Pumpkin    & Blender     &  1 & 1 & No  & Varies     \\ 
    \hline
    Desk       & Blender     &  1 & 3 & No  & Varies     \\ 
    \hline
    Sofa       & Blender     &  4 & 4 & No  & Varies     \\ 
    \hline
    Cake       & Blender     &  4 & 4 & Yes & Varies     \\ 
    \hline
    Spaceship  & Blender     &  4 & 4 & Yes & Fixed     \\ 
    \hline
    Stair      & Blender     &  6 & 4 & Yes & Varies     \\ 
    \hline
    Door       & Blender     &  2 & 1 & No  & Varies     \\ 
    \hline
    Drawer     & Blender     &  4 & 4 & No & Varies     \\ 
    \hline
    Cupcake    & Blender     &  1 & 1 & Yes  & Varies     \\ 
    \hline
    Tree       & Online      &  2 & 7 & No  & Fixed     \\ 
    \hline
    Mussel     & Infinigen   &  2 & 2 & No  & Varies     \\ 
    \hline
    Fork       & Infinigen   &  3 & 2 & No  & Varies     \\ 
    \hline
    Mushroom   & Infinigen   &  2 & 2 & Yes  & Varies     \\ 
    \hline
    Rock       & Infinigen   &  2 & 1 & No  & Varies     \\ 
    \hline
    House      & Blender     &  1 & 4 & Yes & Fixed     \\ 
    \hline
    Chair      & Infinigen   & 10 & 3 & Yes & Varies     \\ 
    \hline
    Clam       & Infinigen   & 2 &  2 & No  & Varies     \\ 
    \hline
    Starfruit  & Infinigen   & 2 &  2 & No  & Varies     \\ 
    \hline
    \bottomrule
    \end{tabular}
    \caption{Procedural Generators Statistics}
    \label{table:Generator Statistics}
\end{table}

\section{ProcCorrNet Training Details}\label{section:appendixB}

One ProcCorrNet for each component group of a given procedural model. For each procedural model, we sample 50 meshes for training. All networks are trained for $10,000$ iterations using the Adam optimizer with a learning rate of $5e^{-5}$, applied across all procedural models used in our experiments and implemented in PyTorch on a single NVIDIA RTX 4090 GPU. 

\section{Author Contributions}
Ruiqi Xu and Zihan Zhu contributed equally to this work. 
Both authors may list their names first on resumes, personal websites, or other professional materials.

\end{document}